\documentclass[aps,onecolumn,pra,showpacs,amsmath,amssymb,showkeys,10pt]{revtex4-2}
\usepackage{graphicx}
\usepackage{epstopdf}
\usepackage{diagbox}
\pdfoutput=1
\usepackage{braket}
\usepackage{hyperref}
\usepackage{longtable}
\usepackage{float}

\begin{document}

	\title{Simulating and investigating various dynamic aspects of the $\rm{H}_2\rm{O}$-related hydrogen bond model}
	
	\author{Jiangchuan You}
	\affiliation{Faculty of Computational Mathematics and Cybernetics, Lomonosov Moscow State University, Vorobyovy Gory 1, Moscow, 119991, Russia}
	
	\author{Ran Chen}
	\affiliation{Faculty of Computational Mathematics and Cybernetics, Lomonosov Moscow State University, Vorobyovy Gory 1, Moscow, 119991, Russia}
	
	\author{Wanshun Li}
	\affiliation{Faculty of Computational Mathematics and Cybernetics, Lomonosov Moscow State University, Vorobyovy Gory 1, Moscow, 119991, Russia}
	
	\author{Hui-hui Miao}
	\email[Correspondence to: Vorobyovy Gory 1, Moscow, 119991, Russia. Email address: ]{hhmiao@cs.msu.ru (H.-H. Miao)}
	\affiliation{Faculty of Computational Mathematics and Cybernetics, Lomonosov Moscow State University, Vorobyovy Gory 1, Moscow, 119991, Russia}

	\author{Yuri Igorevich Ozhigov}
	\email[Correspondence to: Vorobyovy Gory 1, Moscow, 119991, Russia. Email address: ]{ozhigov@cs.msu.ru (Y.I. Ozhigov)}
	\affiliation{Faculty of Computational Mathematics and Cybernetics, Lomonosov Moscow State University, Vorobyovy Gory 1, Moscow, 119991, Russia}

	\date{\today}

	\begin{abstract}
	A basic model of hydrogen bonds related to $\rm{H}_2\rm{O}$, which is adapted from the Jaynes--Cummings model, is suggested, and its different dynamic features are studied theoretically. In this model, the making and breaking of hydrogen bonds happen alongside the creation and destruction of phonons in the surrounding medium. A number of simplifying assumptions about the dynamics of the molecules involved are used. The rotating wave approximation is applied under consideration of the strong-coupling condition. Dissipative dynamics under the Markovian approximation is obtained through solving the quantum master equation --- Lindbladian. We obtain the probabilities of reaction channels involving hydrogen bonds based on the parameters of the external environment. Differences between unitary and dissipative evolutions are discussed. Consideration is given to the effects of all kinds of potential interactions and dissipation on evolution. Consideration is also given to the reverse processes (inflows) of dissipation. The results show that the magnitude changes of the interactions and dissipation have a slight effect on the formation of hydrogen bonds, but the variation of the inflows significantly affects the formation of hydrogen bonds. According to the findings, the dynamics of the $\rm{H}_2\rm{O}$-related hydrogen bond model can be controlled by selectively choosing system parameters. The results will be used as a basis to extend the research to more complex chemical and biological models in the future.
	\end{abstract}


	\maketitle

	\section{Introduction}
	\label{sec:Intro}
	
	Complex quantum system modeling is one of the most important directions in computational mathematics today, especially in the computational fields involving polymer chemistry and macromolecular biology \cite{McArdle2020, Baiardi2023, Albuquerque2021}. A computer chemistry simulator with predictive power requires an entirely quantum treatment; its creation poses a serious challenge to computational mathematics due to the exponentially growing complexity of calculations --- curse of dimensionality \cite{Bellman1957, Bellman1961}. The dynamics of chemical transformations, in contrast to the molecular dynamics of ready-made molecules, require the involvement of an electromagnetic field, which further aggravates the problem of complexity. The most important type of chemical transformation is the formation and disintegration of hydrogen bonds between molecules, which are responsible for the formation and disintegration of macromolecules. Such bonds are formed by a proton tunneling between two conventional potential wells between two molecules. The discovery of hydrogen bonds is attributed to T.S. Moore and T.F. Winmill \cite{Moore1912}, and the description of hydrogen bonding in water was first described in 1920 \cite{Latimer1920}. Hydrogen bonds are much weaker than covalent bonds (in a water ($\rm{H}_2\rm{O}$)$_2$ dimer, the energy of a hydrogen bond is only an order of magnitude higher than the thermal energy at room temperature, while for a covalent bond in an $\rm{OH}$ molecule the energy is 200 times greater), and therefore their formation and decay are easily controlled by external influences; for example, temperature serves as a mechanism for the transformation of macromolecules. Such transformations occur, for example, during the synthesis of DNA, the double helix of which is connected precisely by these bonds. Hydrogen bonds in water are responsible for its extreme heat capacity; their short lifetime is about $10^{-11}$ seconds \cite{Dillon2012} --- determines the flexibility of water clusters and their good interaction with donor molecules \cite{Stahl1986}. In recent years, hydrogen bonds have become one of the main objects of research into quantum processes related to biology. Decoherence in hydrogen bonds was considered in \cite{Ignacio2023}. The entangled spin states that arise in them are in \cite{He2022}. A more chemical consideration of the hydrogen bonds that arise in the $\alpha$-helix of proteins participating in the protein machinery of living organisms is presented in \cite{Danko2022}. Hybrid bonds in liquids, including proton tunneling, as well as in water clusters, were studied in \cite{Di2018, Yamada2020}. The possibility of using proton tunneling to recognize molecules was also explored in \cite{Pusuluk2018}. The chemical role of hydrogen bonding in enzymes was studied in \cite{Farrow2018}.
	
	Consideration of this type of bond involves many elements. In our work, we propose a highly simplified model of hydrogen bonds that can be easily scaled to complex molecular systems, making their simulation possible on modern computers. A key contribution of this paper is the cavity quantum electrodynamics (QED) models \cite{Rabi1936, Rabi1937, Dicke1954, Hopfield1958, Casanova2010}, which are easy to implement in the laboratory and offer a unique scientific paradigm for studying light--matter interaction. The cavity QED model includes the Jaynes--Cummings model (JCM) \cite{Jaynes1963} and the Tavis--Cummings model (TCM) \cite{Tavis1968}, as well as their generalizations \cite{Angelakis2007}. Many studies have been conducted recently in the field of these models, including those on quantum gates \cite{OzhigovYI2020, Dull2021}, quantum many-body phenomena \cite{Smith2021}, entropy \cite{MiaoHuihui2024}, quantum discord \cite{MiaoLi2025}, dark states \cite{Lee1999, Andre2002, Poltl2012, Tanamoto2012, Hansom2014, Kozyrev2016, Ozhigov2020, Afanasyev2022}, phase transitions \cite{Prasad2018, Wei2021}, etc. \cite{Guo2019, Victorova2020, Ozhigov2021, Kulagin2022, Pluzhnikov2022, Chen2022, Miao2023, MiaoOzhigov2023, LiMiao2024, MiaoOzhigov2024}. As a basis, the generally accepted JCM is introduced and modified appropriately so that the presence of a hydrogen bond will play the role of the ground state of (conditional) atoms, and its absence will play the role of the excited states of the atoms. The optical cavity will correspond to the region where the emitted phonon can be again absorbed by the molecular structure through the destruction of the resulting hydrogen bond.
	
	This paper is organized as follows. The $\rm{H}_2\rm{O}$-related hydrogen bond model is proposed in Sec. \ref{sec:HBModel}. After introducing the physico-biological mechanisms of the target model in Sec. \ref{subsec:Target}, its Hilbert space and Hamiltonian are constructed in Sec. \ref{subsec:StatesHamil}, and the quantum master equation (QME) is introduced in Sec. \ref{subsec:QME}. The numerical method is introduced in Sec. \ref{sec:Method}. The results of our numerical simulations are presented in Sec. \ref{sec:Results}, including a comparison between unitary and dissipative evolutions in Sec. \ref{subsec:Comparison}, and the effects of interactions, dissipation, and reverse processes (inflows) of dissipation on the evolution in Secs. \ref{subsec:EffectInter} $\sim$ \ref{subsec:EffectInflu}. Besides, the effect of external impulses on evolution is shown in Sec. \ref{subsec:external_impulses}. Some brief comments on our results in Sec. \ref{sec:Conclusion} close out the paper. Some technical details are included in Appendices \ref{appx:QME} and \ref{appx:PhysQuantities}.
	
	\section{Hydrogen bond model}
	\label{sec:HBModel}

    The hydrogen bond dynamics in the medium are represented as a dynamics of polariton, which consists of a group of real particles (two molecules and photons) and quasi-particles (phonons). The reason why we introduce phonons is that the formation and breaking of hydrogen bonds belong to intermolecular interactions. This intermolecular vibration exists in the form of phonons in solid ice crystals; while for liquid water, although it does not have the long-range order characteristics of ice crystals, it still has short-range order characteristics within 1 to 5 atomic distances. At this time, the concept of local phonons or quasi-phonons can still be used to represent intermolecular vibrations. Strictly speaking, the quantum state of polariton has the following form
    \begin{equation}
    		\label{eq:QuantumState}
    		\lambda_{photon}|photon\rangle+\lambda_{phonon}|phonon\rangle+\lambda_{particle}|particle\rangle.
    \end{equation}
    This is hardly possible to deal with in this form because any interaction (photon--phonon, phonon--atom, photon--atom) represents a nontrivial task. Thus, the initial photon can transform into a few phonons. So it is accepted to call the polariton of phonons if there are no explicit photons in advance. This terminology is followed, and the system in the framework of the Jaynes--Cummings scheme is presented by the Hamiltonian
    \begin{equation}
    		\label{eq:Interation}
    		a^{\dag}\sigma+a\sigma^{\dag}+a^{\dag}a+\sigma^{\dag}\sigma,
    \end{equation}
    where the field operator $a$ relates to our conditional phonon, and operator $\sigma$ --- to real particles.

    The weakness of the hydrogen bond causes its strong dependence on external conditions, in particular, on the ambient temperature. The temperature itself can be conventionally represented as a graph of the average number of phonons versus their frequency. In \cite{Huelga2013}, the temperature effect on atoms within the Jaynes--Cummings model was represented as terms of the Hamiltonian of the form
    \begin{equation}
            g(a_{\omega}^\dag +a_{\omega})\sigma_{\omega}^\dag\sigma_{\omega},
            \label{eq:Hu}
    \end{equation}
    where $a_{\omega}$ is the annihilation operator of the phonon of the selected mode $\omega$, $\sigma_{\omega}$ is the relaxation operator of the atom. This approach is extremely computationally expensive, since it requires explicit inclusion of phonons in the basic states of the model, which immediately causes a huge increase in the required memory and does not allow scaling the model to multi-molecular structures to study new collective effects. In addition, the case of atomic excitations is very different from Eq. \eqref{eq:Hu}, because their energy is several orders of magnitude greater than the energy of a single phonon, so that a term of the form Eq. \eqref{eq:Hu} corresponds only to dephasing, but not to the direct interaction of phonons with matter.
            
    The decoherence is described as caused by the influence of the medium using the QME with a decoherence factor in the form of phonon annihilation. This approach is simpler and more efficient than an independent a priori introduction of decoherence with a Gaussian factor in \cite{Ignacio2023} and also allows for simple scaling.
            
    Interest is only given to the frequency that most strongly affects the hydrogen bond. This influence can be represented as a term of the Hamiltonian of the form 
    \begin{equation}
            H_{int}=g_{hb}(a_{phn}^{\dag}\sigma_{hb}+a_{phn}\sigma_{hb}^\dag),
    \end{equation}
    where $\sigma_{hb}$ is the operator of the hydrogen bond formation accompanied by phonon emission, and $a_{phn}$ is the operator of phonon absorption leading to bond decay. Direct inclusion of such a term in the Hamiltonian is also unacceptable for us, since explicit phonons again arise in the basic states. An approximation in the form of the average number of phonons $n_{av}$ at the resonance frequency sensitive to hydrogen bonding is applied; since the operators of phonon annihilation and creation are proportional to the square root of their number $\sqrt{n_{av}}$, $H_{int}$ is replaced with the operator
    \begin{equation}
	    H'_{int}=g_0\sqrt{n_{av}}(\sigma_{hb}^\dag+\sigma_{hb}).
    \end{equation}
    Then the basic state will not contain an explicit number of phonons, and the model is scalable to more complex molecular systems.
        
    How to relate $n_{av}$ to the temperature of the environment of interacting molecules? The thermally stable state of the phonon field inside the cavity is introduced and has the following form
    \begin{equation}
            \label{stab}
            {\cal G}\left(T\right)=c\sum\limits_{n=0}^\infty e^{\frac{-\hbar\omega n}{KT}}|n\rangle\langle n|,
    \end{equation}
    where $\hbar$ is the reduced Planck constant, $\omega$ is the phononic mode, $n$ is the phonon number, and $T$ is the cavity temperature at a given frequency mode $\omega$, $K$ is the Boltzmann constant, and $c$ is the normalization factor. $\gamma_{in}$ and $\gamma_{out}$, which are the intensities of the inflow and dissipation of phonons into and out of a notional cavity around the molecules, are proposed to form a geometric progression with the denominator \cite{Kulagin2019}
    \begin{equation}
            \label{eq:Ratio}
            \mu=\frac{\gamma^{in}}{\gamma^{out}}=e^{\frac{-\hbar\omega}{KT}},
    \end{equation}
    where $\gamma^{in}$ refers to the overall spontaneous inflow rate and $\gamma^{out}$ refers to the overall spontaneous emission rate. A stable temperature occurs only when $\gamma_{in}<\gamma_{out}$. Knowing these coefficients, or knowing the temperature $T$ directly, $n_{av}$ is obtained. In practice, the coefficients are found only by optimizing them using neural networks based on the experimental results. The temperature at a fixed photonic mode is determined in a similar way. Photonic modes relate to transformations of electron states, and phononic modes relate to proton oscillations.
 
	\subsection{The target model}
    \label{subsec:Target}
    
\begin{figure}
		\centering
        \includegraphics[width=1.\textwidth]{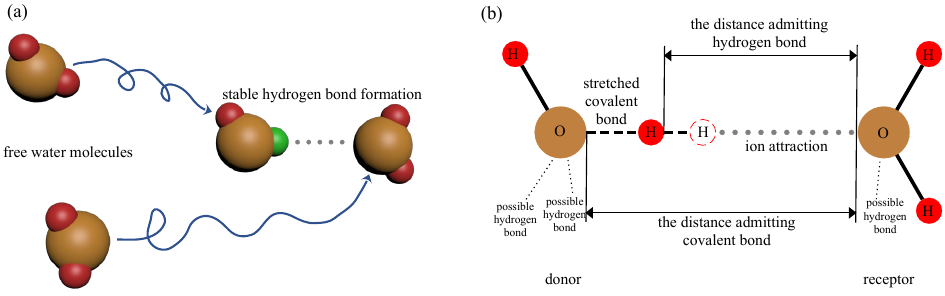}
        \caption{(online color) {\it Hydrogen bond formation mechanism.} A scale model of two $\rm{H}_2\rm{O}$ molecules is shown in panel (a). The mechanism of ion attraction between hydrogen and oxygen atoms is shown in the form of a ball--and--stick model in panel (b). Brown circles (balls) represent oxygen atoms, red circles (balls) represent hydrogen atoms in an excited state, black solid lines stand for covalent bonds, and gray dotted lines represent ion attraction; especially, black dashed lines indicate stretched covalent bonds, and green balls in panel (a) represent protons in a ground state.}
        \label{fig:HBFormation}
	\end{figure}

    Interaction between two water molecules causes micro-oscillations of the hydrogen atom in one of the water molecules; it allows hydrogen bonds to form between water molecules, but this process does not break covalent bonds. The hydrogen bond formation mechanism is as shown in Fig. \ref{fig:HBFormation}. In panel (a), when two $\rm{H}_2\rm{O}$ molecules that are moving freely are far apart, a hydrogen bond cannot form between the molecules. However, when the molecules move closer together, the hydrogen atom (proton) of one molecule and the oxygen atom of another molecule attract each other, and a stable hydrogen bond is formed. In panel (b), the hydrogen atom of one molecule, the “donor,” is attracted by the oxygen atom of another molecule, the “receptor,” to produce the tunneling effect. The proton tunneling in the normal state does not destroy the covalent bond between the proton and its parent molecule but only deforms it. In the case without considering electrons, the effect of some stretching of the covalent bond on the hydrogen bond formation is ignored, and it is considered a normal state with a covalent bond to avoid cluttering the target model. In the model with electrons, this difference is no longer ignored. Theoretically, in addition to 2 covalent bonds for the oxygen atom, there are actually 2 more hydrogen bonds with protons from neighboring molecules. The oxygen atom, and its covalent and hydrogen bonds form a tetrahedron.
    
    A hydrogen bond forms between water molecules when they get close enough, allowing one of the protons attached to one molecule to move between it and a nearby molecule while still keeping its bond. The qubit $|d\rangle_{dist}$ is introduced to describe the relative position of the proton in the system, and $d\in\{ -1,0,1,2\}$. $dist=2$ indicates a significant distance between the molecules, rendering the formation of a hydrogen bond theoretically impossible. The state $dist=1$ means that the molecules are approaching each other at a critical distance, and although it does not allow the formation of a hydrogen bond, it allows the molecules to tunnel to the state $dist=0$, where the bond becomes possible. The state $dist=-1$ means the presence of a stable hydrogen bond; here, it is an irreversible process. However, researchers investigate the influence of temperature on the hydrogen bond formation, as a change in temperature triggers the inflow of phonons. Therefore, in this situation, the process becomes reversible, allowing for the breaking of hydrogen bonds.
    
    A term of the Hamiltonian of the form $H_iCond(A)$ with the condition operator $Cond(A)$ means that if the Hamiltonian $H_i$ has the form
    \begin{equation}
        H_i=H_i^{basis}+H_i^{cogerence},
        \label{eq:EquationHcon}
    \end{equation}
    where $H_i^{basis}$ is diagonal in the standard basis, and $H_i^{cogerence}$ is not diagonal in this basis. The term $H_i^{basis}$ is added to the total Hamiltonian in any case, and $H_i^{cogerence}$ is added only if the condition $A$ is satisfied. This remark about the conditional Hamiltonian will be valid for further models as well. This method achieves equality of rest energies for all terms in the general Hamiltonian. So the condition operator $Cond$ means only the inclusion of coherent terms in the Hamiltonian. For example, the proton (and, subsequently, the electron) jumps only if $d=0$, that is, when the molecules are close. Only at a large distance between the molecules is it as if it is ``frozen'' and does not lead to a change in the state of the proton (and, subsequently, the electron). And as soon as the molecules are close, the motions of these particles begin. Thus, the standard basis distinguishes itself from all other bases. In this work, we consider a single hydrogen bond between a pair of water molecules, which will be extended to more hydrogen bonds using Eq. \eqref{eq:EquationHcon} in the future.

	Hydrogen bonding is a fundamental interaction in chemistry and biology, playing a crucial role in the structure and function of a wide range of molecules, from water and small organic compounds to large biomolecules like DNA and proteins. While there is no universally agreed-upon value for the hydrogen bond formation time, empirical studies suggest it typically occurs within the femtosecond to picosecond range, that is, from $10^{-15}$ to $10^{-12}$ seconds. This is due to the fact that the formation of hydrogen bonds involves the reorganization of electron clouds --- a very fast quantum mechanical process. Time-scale measurements of the hydrogen bond dynamics can be performed using ultrafast spectroscopy techniques \cite{Fecko2003, Lawrence2003, Moller2004}.
    
    \subsection{States and Hamiltonian}
    \label{subsec:StatesHamil}
    
    \begin{figure}
         \centering
        \includegraphics[width=1.\textwidth]{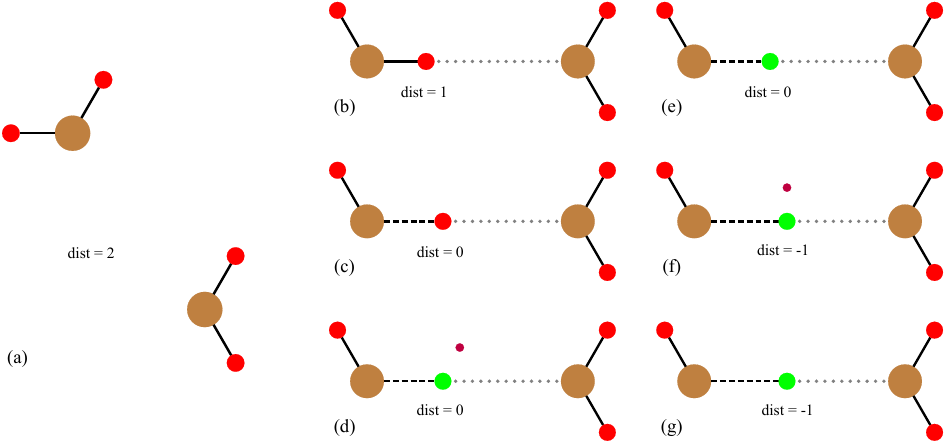} 
        \caption{(online color) {\it Quantum states involved in evolution.} (a) $\sim$ (g) correspond to seven states for the $\rm{H}_2\rm{O}$-related hydrogen bond model: $|2\rangle|1\rangle|0\rangle,\ |1\rangle|1\rangle|0\rangle,\ |0\rangle|1\rangle|0\rangle,\ |0\rangle|0\rangle|1\rangle,\ |0\rangle|0\rangle|0\rangle,\ |-1\rangle|0\rangle|1\rangle,\ |-1\rangle|0\rangle|0\rangle$. Here, brown, red, green, and purple circles represent oxygen atoms, protons with excited states, protons with ground states, and phonons, respectively; solid and dashed black lines stand for ordinary and stretched covalent bonds, respectively; and ion attraction is denoted by gray dotted lines.}
        \label{fig:States}
    \end{figure}
    
    In Sec. \ref{subsec:Target}, the hydrogen bond model of a pair of $\rm{H}_2\rm{O}$ molecules is proposed, and the basic state is represented by the following form
    \begin{equation}
    		|d\rangle_{dist}|p\rangle_{prot}|n\rangle_{phn},
    		\label{eq:DMP}
    \end{equation}
    where the first qubit $|d\rangle_{dist}$ is the relative position of two water molecules, which is detailed in Sec. \ref{subsec:Target}; the second quantum bit $|p\rangle_{prot}$ is the state corresponding to the proton, $p=0$ --- proton is in the ground state $|\Phi_0^{pr}\rangle$ and $p=1$ --- proton is in the excited state $|\Phi_1^{pr}\rangle$; the third qubit $|n\rangle_{phn}$ is the number of phonons corresponding to the hydrogen bond. A special case of our model is proposed, where at most one phonon is pumped into the system (at this time, $n_{av}=1$). When a hydrogen bond is formed, a phonon is emitted. When a hydrogen bond is broken, the phonon is absorbed.
    
	According to Eq. \eqref{eq:DMP}, when $n_{av}=1$, the Hilbert space of the $\rm{H}_2\rm{O}$-related hydrogen bond model will consist of the following 7 basic states, which are shown in Tab. \ref{tab:States}. And the ball--and--stick forms of these states are shown in Fig. \ref{fig:States}.
    \begin{table}[!htpb]
        \centering
	\begin{tabular}{|c|c|}
		\hline
		Qubit & Description \\
		\hline
		$|2\rangle|1\rangle|0\rangle$ & Broken state \\
            \hline
		$|1\rangle|1\rangle|0\rangle$ & Critical state \\
            \hline
            $|0\rangle|1\rangle|0\rangle$ & Stretched and excited state \\
            \hline
            $|0\rangle|0\rangle|1\rangle$ & Stretched and ground state with a phonon \\
            \hline
            $|0\rangle|0\rangle|0\rangle$ & Stretched and ground state without a phonon \\
            \hline
            $|-1\rangle|0\rangle|1\rangle$ & Stable state with a phonon \\
            \hline
            $|-1\rangle|0\rangle|0\rangle$ & Stable state without a phonon \\
		\hline
	\end{tabular}
	\caption{{\it Quantum states involved in evolution.} The evolution involves a total of 7 states. We use the following modifiers, for convenience, to define the different states: “Broken” means the hydrogen bond is broken and two molecules are in a free state; “critical” means the critical point of the hydrogen bond formation and breaking is obtained; “stretched” means the proton of the donor moves toward the receptor due to the ion attraction, and the covalent bond is stretched; “excited” means the proton state is excited; “ground” means the proton state is ground; and “stable” means the hydrogen bond is stable. In particular, a stable state necessitates both stretching and grounding. Besides states $|-1\rangle|0\rangle|1\rangle$ and $|-1\rangle|0\rangle|0\rangle$, the rest of the states can all be defined as “unstable.” In addition, there are two cases for the ground state: with/without a phonon.}
	\label{tab:States}
    \end{table}
    
    $H_{hb}$ is represented by the Hamiltonian of a system of two water molecules connected by a hydrogen bond and has the following form
    \begin{equation}
    	H_{hb} = H_{dist}Cond(p=1) + H_{prot}Cond(d=0),
    	\label{equationH_hyd}
    \end{equation}
    where $H_{dist}$ describes the proton tunneling energy and $H_{prot}$ describes the Hamiltonian for transitions. $Cond(p=1)$ and $Cond(d=0)$ are conditional operators. The rotating wave approximation (RWA) is particularly useful in systems where the interactions are resonant or nearly resonant \cite{Wu2007}. RWA is taken into account, and the Hamiltonian is described in the following form
	\begin{itemize}
		\item $H_{dist}$ is defined as follows
		\begin{equation}
            H_{dist} =  \hbar\omega_{dist}\sigma^{\dag}_{dist} \sigma_{dist}+ g_{dist}\left(\sigma^{\dag}_{dist} + \sigma_{dist} \right),
		\end{equation}
		where $\omega_{dist}$ is the frequency for tunneling of phonons; $g_{dist}$, which is located on the non-diagonal line of the Hamiltonian, is the strength for tunneling of protons. And definitions of operators $\sigma_{dist}$ and $\sigma^{\dag}_{dist}$ are as follows
		\begin{equation}
			\sigma_{dist}|1\rangle_{dist}=|0\rangle_{dist},\ \sigma_{dist}^{\dag}|0\rangle_{dist}=|1\rangle_{dist}.
		\end{equation}
 
    		\item $H_{prot}$ is defined as follows
    		\begin{equation}
            H_{prot} = \hbar\omega_{phn}a^{\dag}_{phn}a_{phn} + \hbar\omega_{prot}\sigma^{\dag}_{prot} \sigma_{prot}+ g_{prot}\left(a_{phn} \sigma^{\dag}_{prot} + a^{\dag}_{phn}\sigma_{prot} \right),
    		\end{equation}
    		where $\omega_{phn}$ is the phononic mode, $\omega_{prot}$ is the mode for transitions of protons, and $\omega_{phn}=\omega_{prot}$; $g_{prot}$ is the strength for transitions of protons. And definitions of operators $a^{\dag}_{phn}\sigma_{prot}$ and $a_{phn}\sigma_{prot}^{\dag}$ are as follows
    		\begin{equation}
			a^{\dag}_{phn}\sigma_{prot}|1\rangle_{prot}|0\rangle_{phn}=|0\rangle_{prot}|1\rangle_{phn},\ a_{phn}\sigma_{prot}^{\dag}|0\rangle_{prot}|1\rangle_{phn}=|1\rangle_{prot}|0\rangle_{phn}.
		\end{equation}    
	\end{itemize}
	
\begin{figure}
        \centering
        \includegraphics[width=.5\textwidth]{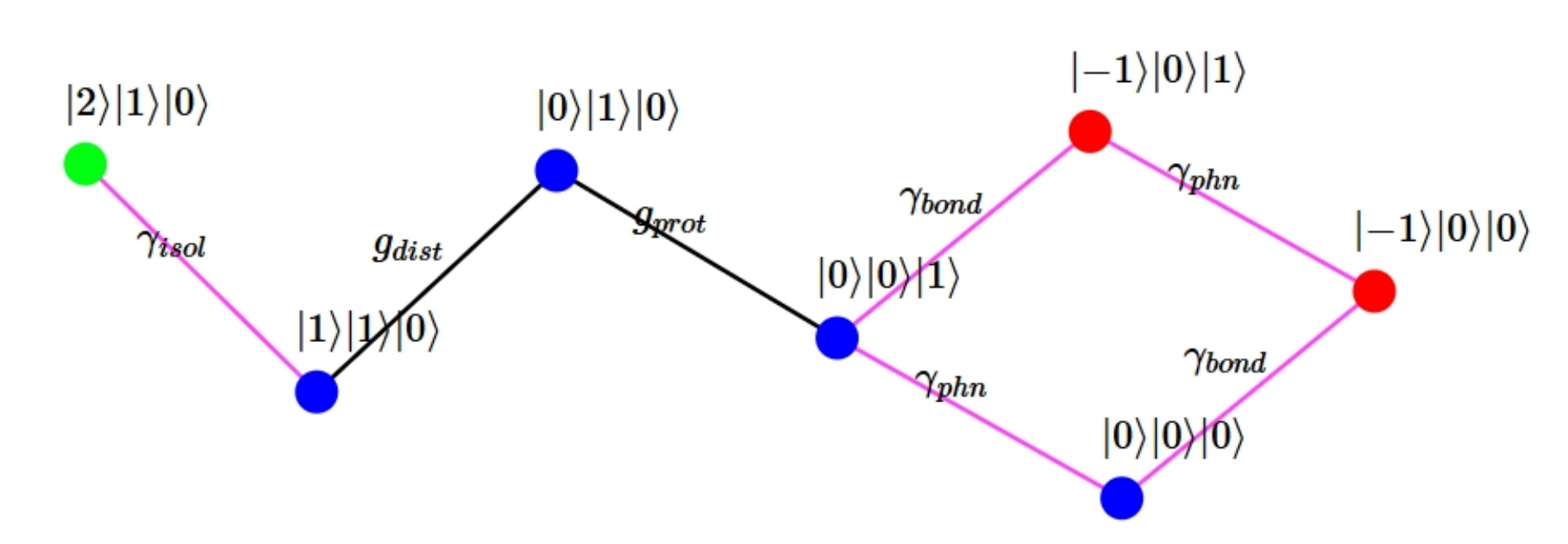}
        \caption{(online color) {\it Network of quantum states and energy wandering.} The quantum system consists of several states, between which there are some potential interactions, dissipation, and inflows. In this figure, the dots represent the states, and the edges represent interactions, dissipation, and inflows. Red dots stand for the states corresponding to the formation of a stable hydrogen bond; the green dot represents the state corresponding to the case that the hydrogen bond is broken and two molecules are free; blue dots represent the states corresponding to an unstable hydrogen bond. Black edges stand for the interactions ($g_{dist}$ and $g_{prot}$); purple edges represent the dissipation and inflows. In particular, $\gamma_{isol}$ includes both dissipation $\gamma_{isol}^{out}$ from state $|1\rangle|1\rangle|0\rangle$ to state $|2\rangle|1\rangle|0\rangle$ and inflow $\gamma_{isol}^{in}$ from state $|2\rangle|1\rangle|0\rangle$ to state $|1\rangle|1\rangle|0\rangle$.}
        \label{fig:Network}
    \end{figure}

	 The complete quantum system ultimately constructed by the total Hamiltonian $H_{hb}$ can be intuitively represented in the form of a network, which is shown in Fig. \ref{fig:Network}. The interactions, dissipation, and inflows between the states can be intuitively seen in this figure, and the energy wandering of the whole system is determined clearly.
	 
     \subsection{Quantum master equation}
     \label{subsec:QME}

    The influence of temperature on the formation of a hydrogen bond requires the use of a QME and the introduction of one decoherence factor related to the qubit $|n\rangle_{phn}$ of the basic state
    \begin{equation}
        	A_{phn}|1\rangle_{phn}=|0\rangle_{phn},\ A_{phn}^{\dag}|0\rangle_{phn}=|1\rangle_{phn}.
        \label{eq:EquationAphn}
    \end{equation}
    And $\gamma_{phn}$ is the dissipation intensity for the escape of the phonons from the cavity to the external environment. Similarly, the two quasi-decoherence factors, which are related to the qubit $|d\rangle_{dist}$ of the basic state, are defined as follows
    \begin{equation}
    		A_{bond}|0\rangle_{dist}=|-1\rangle_{dist}, \ A_{bond}^{\dag}|-1\rangle_{dist}=|0\rangle_{dist}.
        \label{eq:EquationAbond}
    \end{equation}
    \begin{equation}
    		A_{isol}|1\rangle_{dist}=|2\rangle_{dist}, \ A_{isol}^{\dag}|2\rangle_{dist}=|1\rangle_{dist}.
        \label{eq:EquationAisol}
    \end{equation}
    And $\gamma_{bond}$ is the dissipation intensity for the micromotions of molecules that must calm down to form a bond, $\gamma_{isol}$ is the dissipation intensity for the macromotions of molecules in the medium.
    
    The QME, which is also called Lindbladian, is used to describe the time-dependent evolution of the density matrix of an open quantum system, which allows us to obtain the dependence of the probability of the hydrogen bond formation on temperature. It can be seen as a natural result of the coupling between the system and the environment under the lowest-order perturbation and Markov approximation
    \begin{equation}
         i\hbar\dot{\rho} =  [H, \rho] + i\mathcal{L}(\rho),
         \label{equation:QME}
    \end{equation}
    where $\mathcal{L}=L^{in} + L^{out}$. $L^{in}$ describes the inflow process, and $L^{out}$ describes the dissipation process. Then
    \begin{equation}
    		\label{eq:Linblad_in_and_out}
        L^{in}(\rho)=\gamma^{in}\left( A^\dag\rho A - \frac{1}{2}\left(  A A^\dag \rho + \rho A A^\dag  \right) \right),\ L^{out}(\rho)=\gamma^{out}\left( A\rho A^\dag - \frac{1}{2}\left( A^\dag A\rho + \rho A^\dag A  \right) \right),
    \end{equation}
    where $A$ and $A^\dag$ can be replaced by $A_{bond},\ A_{isol},\ A_{phn},\ A_{bond}^\dag,\ A_{isol}^\dag$ and $A_{phn}^\dag$. The full form of the operator $\mathcal{L}$ with consideration of all types of inflows and dissipation is shown in Appx. \ref{appx:QME}.
  
    \section{Numerical method}
    \label{sec:Method}
    
    We use various numerical methods, such as the Euler method, to solve the QME. The iteration consists of three steps. The first step is to calculate the unitary evolution of the density matrix
    \begin{equation}
		\widetilde{\rho}(t+\Delta t) = e^{\frac{-iH\Delta t}{\hbar}} \rho(t) e^{\frac{iH\Delta t}{\hbar}},
		\label{equation:Eulermethod1}
	\end{equation}
    where $\Delta t$ is the iteration time step, and the second step is to apply the Lindblad superoperators \eqref{eq:Linblad_in_and_out} to the density matrix $\rho(t+\Delta t)$ at this moment
    \begin{equation}
		\rho(t+\Delta t) = \widetilde{\rho}(t+\Delta t) + \frac{1}{\hbar} \mathcal{L}(\widetilde{\rho}(t+\Delta t))\Delta t.
		\label{equation:Eulermethod2}
	\end{equation}
    The third step is to scale the density matrix to maintain its positive definiteness, hermiticity, normalization, and other properties.
    
    In addition to the Euler method, there is also the more accurate Runge-Kutta method. For quantum systems with a particularly large number of states, the Monte Carlo wave function method \cite{Dum1992} or the pure state vector method \cite{You2023} can achieve higher computational efficiency. The $\rm{H}_2\rm{O}$-related hydrogen bond model contains a total of 7 states, so using the Euler method makes it easier for us to study the effects of various interactions and dissipation on evolution.

    When solving the QME using the Euler method, the order of magnitude of the time step $\Delta t$ usually depends on the unit of the system's Hamiltonian and the units of other related parameters. The measurement time, which is around 1 $\mu s$, determines the time step in the optical cavity QED experiments with high-finesse optical resonators \cite{Hennrich2000} or microwave resonators \cite{Raimond2001}. However, for a single hydrogen bond, the time step reference measurement time of 1 $\mu s$ is not appropriate. Therefore, the characteristic time scale $\tau$ of the simulated dissipative quantum system is adopted, which can be estimated from the system energy and the reduced Planck constant $\tau =\frac{\hbar}{E}$, where $E$ is the energy of the quantum system and the intensity factor $\gamma$ is usually smaller than $E$. When implementing the Euler method numerically, the time step $\Delta t$ should be significantly smaller than the characteristic time scale.
    
    The energy of a single hydrogen bond $E_{H-bond}$ can be estimated from the average energy of hydrogen bonds in water and Avogadro's constant, which is approximately $0.217655\ eV$. Using the joule as the energy unit may cause computational errors due to extremely small parameter values, so $eV$ is used as the energy unit. And the characteristic time scale $\tau$ is about $3.175316 \times 10^{-15}\ s$. The detailed calculation process is shown in Appx. \ref{appx:PhysQuantities}.
    
    \section{Simulations and results}
    \label{sec:Results}
    
    For convenience, the Hilbert space is divided into two subspaces: the stable hydrogen bond subspace and the unstable hydrogen bond subspace. The stable subspace includes the states $|-1\rangle|0\rangle|1\rangle$ and $|-1\rangle|0\rangle|0\rangle$, and the unstable subspace includes the rest. The stable subspace can be defined as follows
    \begin{equation}
        \label{eq:StableSubspace}
        |\rm{O}\cdots \rm{H}\rangle=c'|-1\rangle|0\rangle|1\rangle+c''|-1\rangle|0\rangle|0\rangle,
    \end{equation}
    where $c',\ c''$ are the normalization factors.

    The physical determination of the time of the hydrogen bond formation involves a combination of experimental techniques such as ultrafast infrared spectroscopy \cite{Lawrence2003}, which determines the time between $0.5ps$ and $1ps$ ($10^{-12}s$). Therefore, in all simulations, two reference parameters are proposed: $g=2\times 10^{-3}\rm{eV},\ \gamma=5\times 10^{-3}\rm{eV}$. All interaction strengths $g_{dist}$ and $g_{prot}$ below are expressed in terms of reference parameter $g$, and all dissipation intensities $\gamma_{isol}^{out}$, $\gamma_{bond}^{out}$, and $\gamma_{phn}^{out}$ below are expressed in terms of reference parameter $\gamma$. Especially, inflow intensities $\gamma_{isol}^{in}$, $\gamma_{bond}^{in}$, and $\gamma_{phn}^{in}$ are not used directly in simulations, but ratios $\mu_{isol}$, $\mu_{bond}$, and $\mu_{phn}$ (inflow intensity divided by dissipation intensity), defined in Eq. \eqref{eq:Ratio}, are used instead. Except for the above-mentioned parameters, which will change during the simulation, the remaining parameters are all fixed. Especially, $\hbar\omega_{prot}+\hbar\omega_{dist} = E_{H-bond}$, $\Delta t = 0.01 \tau$.

    Various dynamic aspects of the $\rm{H}_2\rm{O}$-related hydrogen bond model are shown in the five subsections below. The supercomputer resources \cite{Voevodin2019} are used to obtain the results of simulations.

    \subsection{Comparison between unitary and dissipative evolutions}
    \label{subsec:Comparison}
    
     \begin{figure}
        \centering
        \includegraphics[width=.45\textwidth]{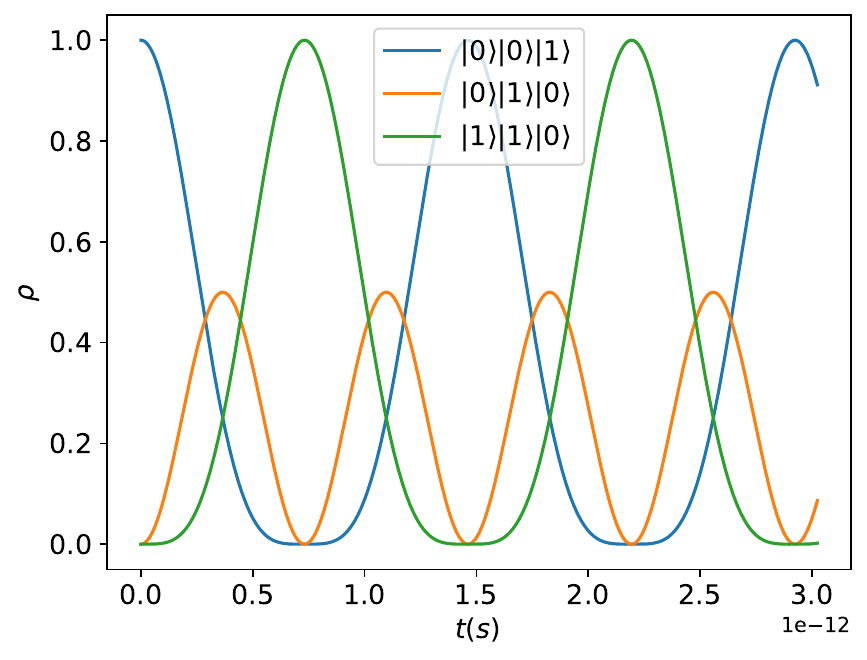}
        \caption{(online color) {\it The unitary evolution.} The initial state is $|0\rangle|0\rangle|1\rangle$. $g_{prot} = g_{dist} = g$ and $\gamma_{isol}^{out}=\gamma_{bond}^{out}=\gamma_{phn}^{out}=0$.}
        \label{fig:Unitary}
    \end{figure}
    
    Firstly, in a closed system without the dissipation intensities, the unitary evolution with only three basic states ($|0\rangle|0\rangle|1\rangle$, $|0\rangle|1\rangle|0\rangle$, and $|1\rangle|1\rangle|0\rangle$) is obtained. The unitary evolution considering only the interaction between particles and fields is calculated by the Schrödinger equation
    \begin{equation}
         i\hbar\dot{\rho} = [H, \rho] = H\rho - \rho H. 
        \label{equation:Sch}
    \end{equation}
    For the unitary evolution, the calculation method only needs to adopt Eq. \eqref{equation:Eulermethod1}. The initial state is $|0\rangle|0\rangle|1\rangle$ --- a stretched and ground state with a phonon. The unitary evolution, considering $g_{prot}=g_{dist}=g$, is shown in Fig. \ref{fig:Unitary}. The three obtained curves, representing the states $|0\rangle|0\rangle|1\rangle$, $|0\rangle|1\rangle|0\rangle$ and $|1\rangle|1\rangle|0\rangle$, oscillate periodically.
    
    \begin{figure}
        \centering
        \includegraphics[width=1.\textwidth]{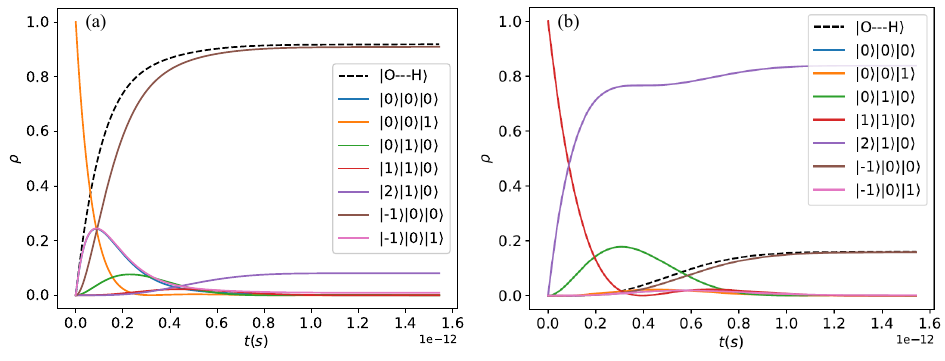}
        \caption{(online color) {\it The dissipative evolution.} $g_{prot} = g_{dist} = g$ and $\gamma_{isol}^{out}=\gamma_{bond}^{out}=\gamma_{phn}^{out}=\gamma$. In panel (a), the initial state is $|0\rangle|0\rangle|1\rangle$; in panel (b), the initial state is $|1\rangle|1\rangle|0\rangle$. The black dashed curve corresponds to the stable hydrogen bond: the probability of $|\rm{O}\cdots\rm{H}\rangle$ is the sum of probabilities of $|-1\rangle|0\rangle|1\rangle$ and $|-1\rangle|0\rangle|0\rangle$.}
        \label{fig:Dissipative}
    \end{figure}
    
    The dissipation intensities are taken into consideration to calculate the dissipative evolution of an open system with phonons entering and exiting. The QME is numerically calculated using Eq. \eqref{equation:Eulermethod2}. In addition to the initial state $|0\rangle|0\rangle|1\rangle$, the initial state $|1\rangle|1\rangle|0\rangle$ --- critical state, is also considered. The critical state means the critical point of the hydrogen bond breaking is reached. A comparison between these two initial states is carried out in Fig. \ref{fig:Dissipative}. As shown in both panels, the periodic oscillations disappear, replaced by the dissipation. The time of the hydrogen bond formation is about $1\rm{ps}$ ($10^{-12}s$) before the probability reaches a stable level. In panel (a), $\rho_{|-1\rangle|0\rangle|0\rangle\langle-1|\langle0|\langle0|} > \rho_{|2\rangle|1\rangle|0\rangle\langle2|\langle1|\langle0|}$, thus, the system tends to form a stable hydrogen bond when the initial state is $|0\rangle|0\rangle|1\rangle$. On the contrary, in panel (b), $\rho_{|-1\rangle|0\rangle|0\rangle\langle-1|\langle0|\langle0|} < \rho_{|2\rangle|1\rangle|0\rangle\langle2|\langle1|\langle0|}$, the system tends to break the hydrogen bond when the initial state is $|1\rangle|1\rangle|0\rangle$. This is because the state $|0\rangle|0\rangle|1\rangle$ is stretched and ground, and the proton of the donor at this time is at a lower energy level. Thus, it is more inclined to form a stable hydrogen bond. The state $|1\rangle|1\rangle|0\rangle$ means that the critical point of the hydrogen bond formation and breaking is obtained, and the hydrogen bond breaks more easily to release two free water molecules.

    \subsection{The effect of interactions on evolution}
    \label{subsec:EffectInter}
    
    \begin{figure}
        \centering
        \includegraphics[width=0.8\textwidth]{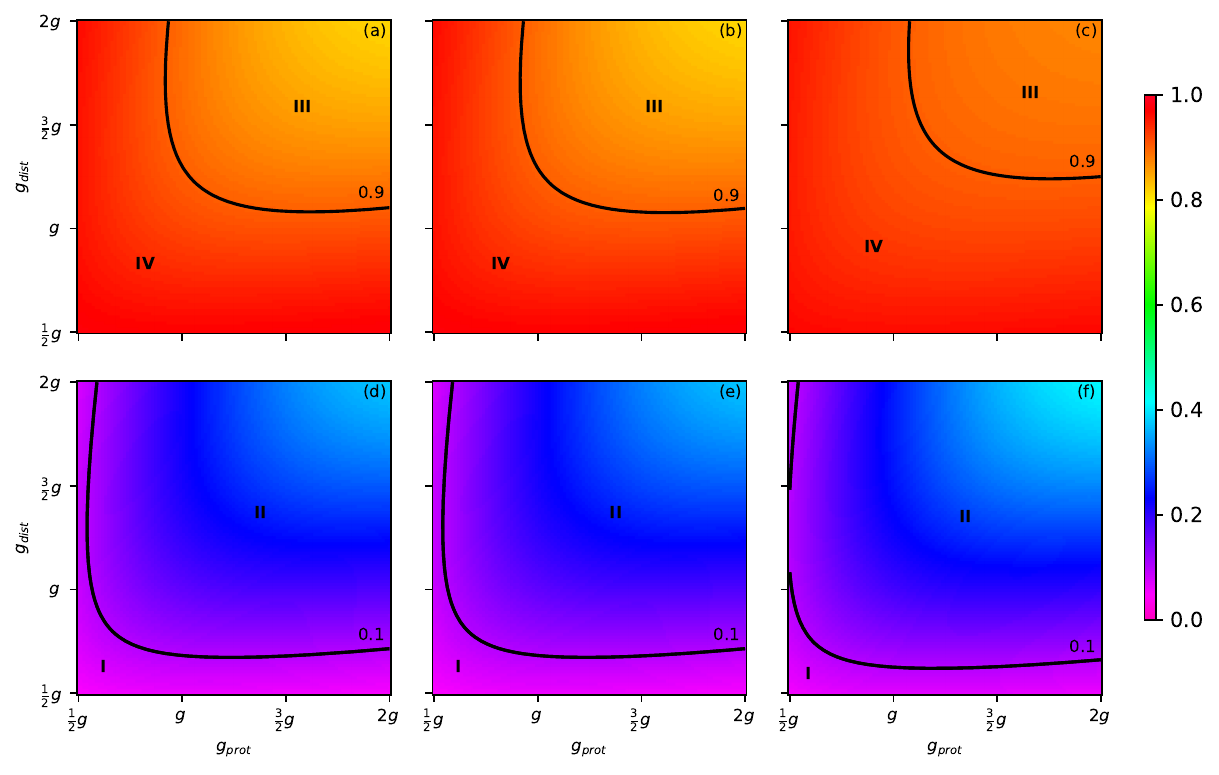}
        \caption{(online color) {\it Effect of different interaction strengths $g_{dist}$ and $g_{prot}$ on the hydrogen bond formation.} The first row of panels is the dissipative evolution result of the initial state $|0\rangle|0\rangle|1\rangle$. The second row of panels is the dissipative evolution result of the initial state $|1\rangle|1\rangle|0\rangle$. The values of $g_{dist}$ and $g_{prot}$ range from $0.5g$ to $2g$, and the intensities of all types of dissipation are fixed: $\gamma_{isol}^{out}=\gamma_{bond}^{out}=\gamma_{phn}^{out}=\gamma$. The first column corresponds to the case that $\mu_{isol}=\mu_{bond}=\mu_{phn}=0$; the second column corresponds to the case that $\mu_{isol}=\mu_{bond}=0,\ \mu_{phn}=0.01$; the third column corresponds to the case that $\mu_{isol}=\mu_{bond}=\mu_{phn}=0.01$. The color of the heat map represents the probability of the stable hydrogen bond formation (the probability of $|\rm{O}\cdots\rm{H}\rangle$ when the system reaches a stable point after a long time). Regions I, II, III, and IV are the regions of probability within $[0,\ 0.1),\ [0.1,\ 0.5),\ [0.5,\ 0.9)$, and $[0.9,\ 1]$, respectively. These definitions of regions I$\sim$IV also apply to Figs. \ref{fig:EffectDissipations} and \ref{fig:EffectInflows}.}
        \label{fig:EffectInteractions}
    \end{figure}
    
	The result of Fig. \ref{fig:EffectInteractions} is shown as heat maps, where the colored dots represent the probability of the stable hydrogen bond formation at $t\to\infty$. Here, $\infty$ is obtained when the system tends to stabilize and the probability of the states approach constant values. In the next subsections, the results of Figs. \ref{fig:EffectDissipations}, \ref{fig:EffectInflows}, and \ref{fig:mu+num} are also represented as heat maps at $t\to\infty$.
	
	The effect of interaction strengths $g_{dist}$ and $g_{prot}$ on the evolution and the hydrogen bond formation is obtained in Fig. \ref{fig:EffectInteractions}, where the values of $g_{dist}$ and $g_{prot}$ range from $0.5g$ to $2g$. All dissipation intensities are fixed: $\gamma_{isol}^{out}=\gamma_{bond}^{out}=\gamma_{phn}^{out}=\gamma$. Under different initial states, the dissipative evolution will eventually reach probabilistic stability after long-term evolution.
    
    In the first row of Fig. \ref{fig:EffectInteractions}, when the initial state is $|0\rangle|0\rangle |1\rangle$, the result demonstrates that the larger the interaction strengths $g_{dist}$ and $g_{prot}$, the lower the probability of the stable hydrogen bond formation. However, in the second row, when the initial state is $|1\rangle|1\rangle |0\rangle$, the opposite result is found. In summary, in the case of $|0\rangle|0\rangle|1\rangle$, the interaction strengths have a negative effect on the formation of the stable hydrogen bond --- hindering the hydrogen bond formation; in the case of $|1\rangle|1\rangle|0\rangle$, the interaction strengths have a positive effect on the formation of the stable hydrogen bond --- promoting the hydrogen bond formation. Besides, when the initial state is $|0\rangle|0\rangle|1\rangle$, the probabilities on the heat maps are all in the range of $[0.5,\ 1]$; when the initial state is $|1\rangle|1\rangle|0\rangle$, the probabilities on the heat maps are all in the range of $[0,\ 0.5)$. This is because, according to the results in Fig. \ref{fig:Dissipative} (a), when the initial state is $|0\rangle|0\rangle|1\rangle$, the probability of the $|\rm{O}\cdots \rm{H}\rangle$ is greater than $0.8$, and in the first row of Fig. \ref{fig:EffectInteractions} the interaction strengths have a slight effect on the evolution, so the probabilities represented by the points on the first row of the heat maps are almost all around $0.8$. Similarly, according to Fig. \ref{fig:Dissipative} (b), when the initial state is $|1\rangle|1\rangle|0\rangle$, the probability of the $|\rm{O}\cdots \rm{H}\rangle$ is less than $0.2$, and in the second row of Fig. \ref{fig:EffectInteractions} the interaction strengths also have a slight effect on the evolution, so the probabilities represented by the points on the second row of the heat maps are almost all around $0.2$.
    
    In this section, the effects of the inflow intensities $\gamma_{phn}^{in}$, $\gamma_{bond}^{in}$, and $\gamma_{isol}^{in}$ are investigated. The first column corresponds to the case that inflows are prohibited: $\mu_{isol}=\mu_{bond}=\mu_{phn}=0$, that is to say, $\gamma_{isol}^{in}=\gamma_{bond}^{in}=\gamma_{phn}^{in}=0$. The second column corresponds to the case that only phonon inflow is permitted: $\mu_{isol}=\mu_{bond}=0,\ \mu_{phn}=0.01$. The third column corresponds to the case that all inflows are permitted: $\mu_{isol}=\mu_{bond}=\mu_{phn}=0.01$. Comparing the second and the third columns with the first column, the result demonstrates that the addition of weak inflows actually promotes the formation of the hydrogen bond, regardless of whether the initial state is $|0\rangle|0\rangle|1\rangle$ or $|1\rangle|1\rangle|0\rangle$. For easy observation, we use the dividing lines. In the first row, from panel (a) to panel (c), it is easy to find that the solid dividing line representing the probability of $0.9$ is shifted to the upper right. At the same time, the area of region III of probability with $[0.5,\ 0.9)$ decreases and the area of region IV of probability with $[0.9,\ 1]$ increases. This means the weak inflows promote the hydrogen bond formation when the initial state is $|0\rangle|0\rangle|1\rangle$. Similarly, in the second row from panel (d) to panel (f), the solid dividing line representing the probability of $0.1$ is shifted to the lower left. At the same time, the area of region I of probability with $[0,\ 0.1)$ decreases, and the area of region II of probability with $[0.1,\ 0.5]$ increases. This means the weak inflows also promote the hydrogen bond formation when the initial state is $|1\rangle|1\rangle|0\rangle$.
    
    In addition to the above results, the effect of different interaction strengths on the time required for the system to reach 
stability is also obtained: the smaller the interaction strengths, the longer it takes to reach stability.
 
    \subsection{The effects of dissipation on evolution}
    \label{subsec:EffectDissi}
    
    \begin{figure}
        \centering
        \includegraphics[width=1.0\textwidth]{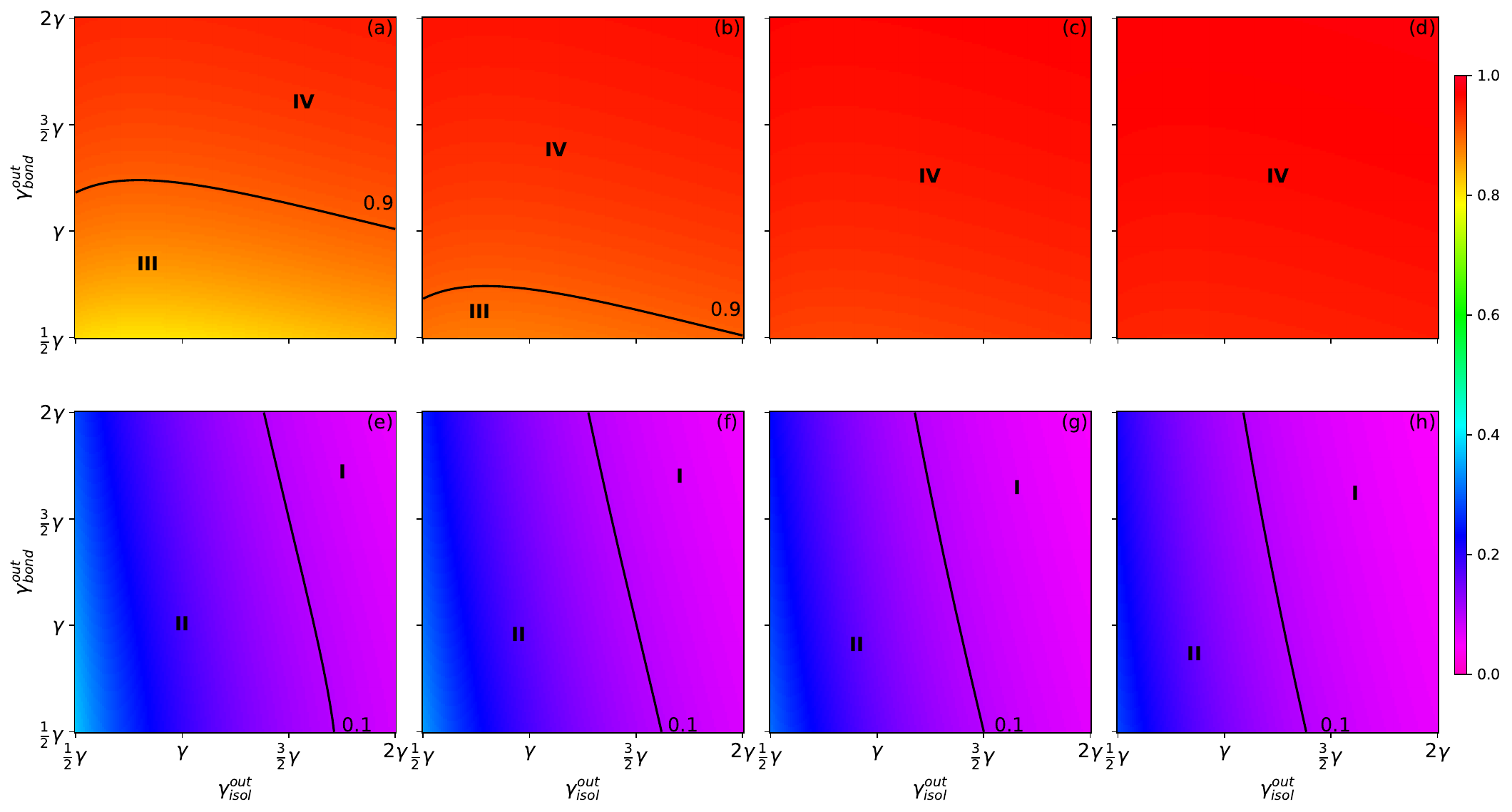}
        \caption{(online color) {\it Effect of different dissipation intensities $\gamma_{isol}^{out}$ and $\gamma_{bond}^{out}$, and $\gamma_{phn}^{out}$ on the hydrogen bond formation.} $g_{prot}=g_{dist}=g$. The first row of panels is the result of the initial state $|0\rangle|0\rangle|1\rangle$. The second row of panels is the result of the initial state $|1\rangle|1\rangle|0\rangle$. The values of $\gamma_{isol}^{out}$ and $\gamma_{bond}^{out}$ range from $0.5\gamma$ to $2\gamma$, and the values of $\gamma_{phn}^{out}$ from the first column to the fourth column correspond to $\frac{1}{2}\gamma,\ \gamma,\ \frac{3}{2}\gamma$, and $2\gamma$, respectively. The definitions of regions II, III, and IV are all defined in Fig. \ref{fig:EffectInteractions}.}
        \label{fig:EffectDissipations}
    \end{figure}
    
    The effect of dissipation on evolution is studied. The parameters being studied are $\gamma_{bond}^{out},\ \gamma_{phn}^{out},\ \gamma_{isol}^{out}$. All interaction strengths are fixed: $g_{prot}=g_{dist}=g$. The values of $\gamma_{isol}^{out}$ and $\gamma_{bond}^{out}$ range from $0.5\gamma$ to $2\gamma$. The values of $\gamma_{phn}^{out}$ from the first column to the fourth column correspond to $0.5\gamma,\ \gamma,\ 1.5\gamma$, and $2\gamma$, respectively. Under different initial states, the dissipative evolution will also eventually reach probabilistic stability after long-term evolution.
    
    In the first row of Fig. \ref{fig:EffectDissipations}, when the initial state is $|0\rangle|0\rangle|1\rangle$, the result demonstrates that the larger the dissipation intensities $\gamma_{isol}^{out}$ and $\gamma_{bond}^{out}$, the higher the probability of forming the stable hydrogen bond. However, in the second row, when the initial state is $|1\rangle|1\rangle|0\rangle$, the opposite result is found. In the first row, from panel (a) to panel (d), it is easy to find that the solid dividing line representing the probability of $0.9$ is shifted from top to bottom until it disappears. At the same time, the area of region III of probability with $[0.5,\ 0.9)$ decreases to $0$, and the area of region IV of probability with $[0.9,\ 1]$ increases to $1$. This means the intensity $\gamma_{phn}^{out}$ promotes the hydrogen bond formation when the initial state is $|0\rangle|0\rangle|1\rangle$. Similarly, in the second row from panel (e) to panel (h), the solid dividing line representing the probability of $0.1$ is shifted from right to left. At the same time, the area of region I of probability with $[0,\ 0.1)$ increases, and the area of region II of probability with $[0.1,\ 0.5]$ decreases. This means intensity $\gamma_{phn}^{out}$ hinders the hydrogen bond formation when the initial state is $|1\rangle|1\rangle|0\rangle$. In summary, in the case of $|0\rangle|0\rangle|1\rangle$, all dissipation intensities have a positive effect on the formation of the stable hydrogen bond; in the case of $|1\rangle|1\rangle|0\rangle$, all dissipation intensities have a negative effect on the formation of the stable hydrogen bond. Similar to Fig. \ref{fig:EffectInteractions}, when the initial state is $|0\rangle|0\rangle|1\rangle$, the probabilities on the heat maps are all in the range of $[0.5,\ 1]$; when the initial state is $|1\rangle|1\rangle|0\rangle$, the probabilities on the heat maps are all in the range of $[0,\ 0.5)$. This shows that the dissipation intensities have a slight effect on evolution, too.
	
    \subsection{The effects of inflows on evolution}
    \label{subsec:EffectInflu}
    
    \begin{figure}
        \centering
        \includegraphics[width=1.0\textwidth]{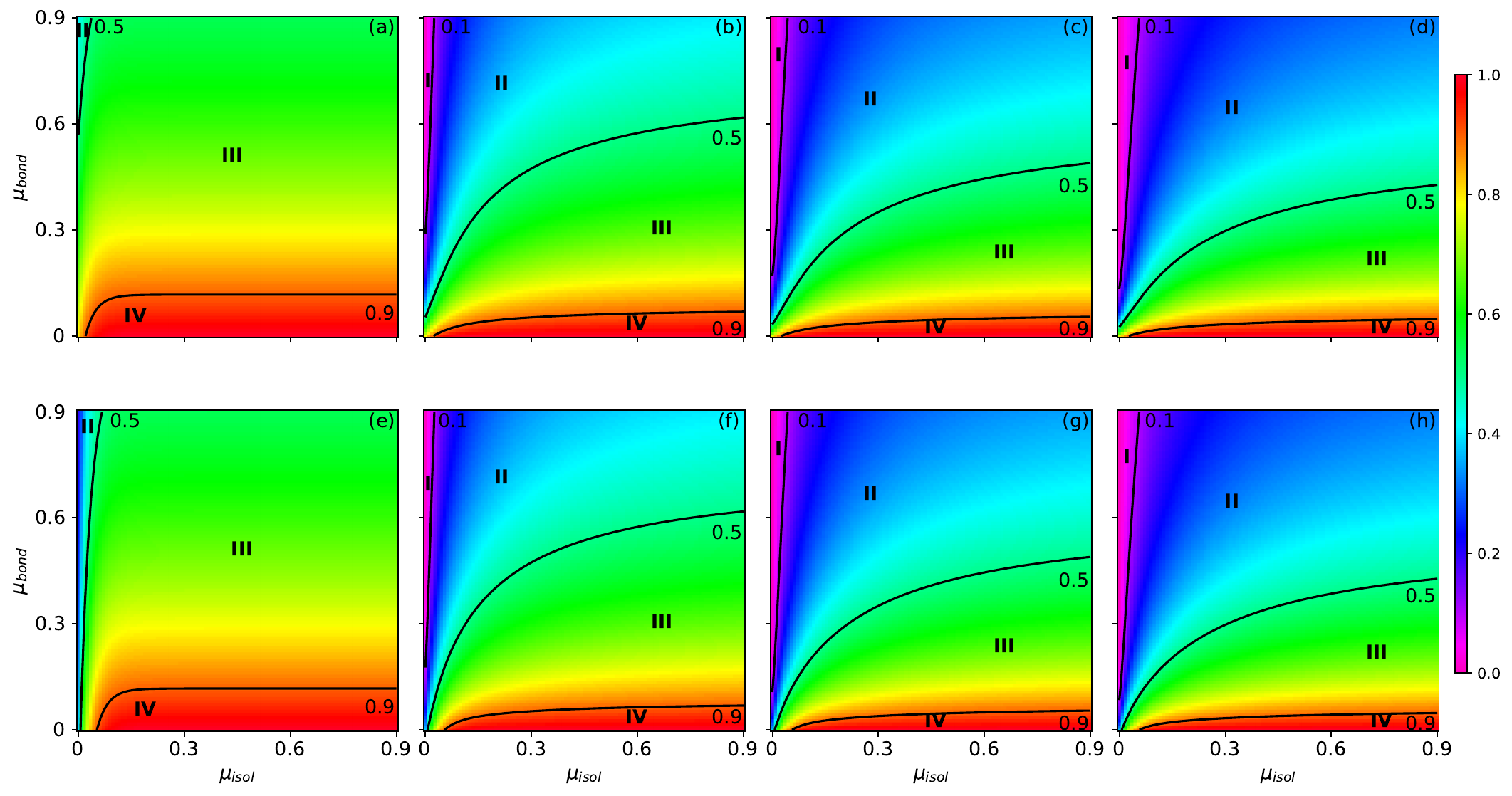}
        \caption{(online color) {\it Effect of different inflow intensities $\mu_{isol}$, $\mu_{bond}$, and $\mu_{phn}$ on the hydrogen bond formation.} $g_{prot}=g_{dist}=g$, $\gamma_{isol}^{out}=\gamma_{bond}^{out}=\gamma_{phn}^{out}=\gamma$. The first row of panels is the result of the initial state $|0\rangle|0\rangle|1\rangle$. The second row of panels is the result of the initial state $|1\rangle|1\rangle|0\rangle$. The values of $\mu_{isol}$ and $\mu_{bond}$ range from $0$ to $0.9$, and the values of $\mu_{phn}$ from the first column to the fourth column correspond to $0,\ 0.3,\ 0.6$, and $0.9$, respectively. The definitions of regions I, II, III, and IV are all shown in Fig. \ref{fig:EffectInteractions}.}
        \label{fig:EffectInflows}
    \end{figure}
    
    Temperature is also an important factor. According to Eq. \eqref{eq:Ratio}, $\mu_{bond},\ \mu_{isol}$, and $\mu_{phn}$ are described as follows
    \begin{equation}
    		\mu_{isol}=\frac{\gamma_{isol}^{in}}{\gamma_{isol}^{out}}=e^{\frac{-\hbar\omega_{isol}}{KT_{isol}}},\ \mu_{bond}=\frac{\gamma_{bond}^{in}}{\gamma_{bond}^{out}}=e^{\frac{-\hbar\omega_{bond}}{KT_{bond}}},\ \mu_{phn}=\frac{\gamma_{phn}^{in}}{\gamma_{phn}^{out}}=e^{\frac{-\hbar\omega_{phn}}{KT_{phn}}},
	\end{equation}
	where $\omega_{isol}=\omega_{bond}=\omega_{phn}$, $\mu_{isol}<1,\ \mu_{bond}<1,\ \mu_{phn}<1$. When $\gamma_{isol}^{out}=\gamma_{bond}^{out}=\gamma_{phn}^{out}=\gamma$, $\mu_{bond},\ \mu_{isol}$ and $\mu_{phn}$ are positively correlated with inflow intensities $\gamma_{isol}^{in}$, $\gamma_{bond}^{in}$ and $\gamma_{phn}^{in}$, respectively; they are also positively correlated with temperature $T_{isol}$, $T_{bond}$ and $T_{phn}$, respectively. For the convenience of research, $\mu_{isol}$, $\mu_{bond}$, and $\mu_{phn}$ can be used to study the effects of inflows or temperatures on evolution. The values of $\mu_{isol}$ and $\mu_{bond}$ range from $0$ to $0.9$. The values of $\mu_{phn}$ from the first column to the fourth column correspond to $0,\ 0.3,\ 0.6$, and $0.9$, respectively. Under different initial states, the dissipative evolution will also eventually reach probabilistic stability after long-term evolution.
	
	According to Fig. \ref{fig:EffectInflows}, the result demonstrates that no matter if the initial state is $|0\rangle|0\rangle|1\rangle$ or $|1\rangle|1\rangle|0\rangle$, the larger $\mu_{isol}$ and the smaller $\mu_{bond}$, the higher the probability of the stable hydrogen bond formation. This indicates that whereas $\mu_{bond}$ inhibits the creation of hydrogen bonds, $\mu_{isol}$ encourages them. In the first row, from panel (a) to panel (d), it is easy to find that all three solid dividing lines representing the probabilities of $0.1$, $0.5$, and $0.9$, respectively, are shifted from left to right. Simultaneously, the areas under the probability curve corresponding to regions I $[0,\ 0.1)$ and II $[0.1,\ 0.5)$ increase, whereas those for regions III $[0.5,\ 0.9)$ and IV $[0.9,\ 1]$ decrease. This means the $\mu_{phn}$ hinders the hydrogen bond formation no matter the initial state is $|0\rangle|0\rangle|1\rangle$ or $|1\rangle|1\rangle|0\rangle$. In summary, the $\mu_{isol}$ has a positive effect on the formation of the stable hydrogen bond; however, the $\mu_{bond}$ and $\mu_{phn}$ have a negative effect on it. Different from Figs. \ref{fig:EffectInteractions} and \ref{fig:EffectDissipations}, the probabilities on the heat maps are all in the range of $[0,\ 1]$. This shows that inflow intensities (temperature) have a more significant effect on evolution than that of other factors (interactions and dissipation).

	\subsection{The effects of external impulses on evolution}		\label{subsec:external_impulses}
	
	\begin{figure}
		\centering
    		\includegraphics[width=1.\textwidth]{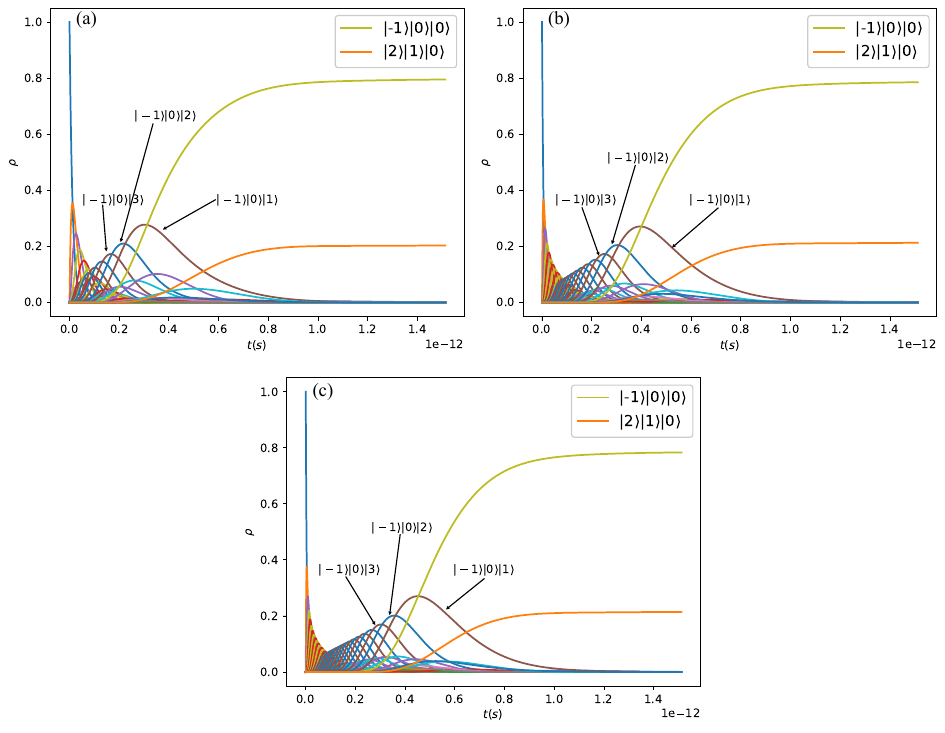}
  		\caption{(online color) {\it Effect of external impulses of phonons on the hydrogen bond formation.} $g_{dist}=g_{prot}=g$, $\gamma_{isol}^{out}=\gamma_{bond}^{out}=\gamma_{phn}^{out}=\gamma$, $\mu_{isol}=\mu_{bond}=0$, $\mu_{phn}=0.01$. The number of phonons varies in different panels: 10 in panel (a), 20 in (b), and 30 in (c).}
		\label{fig:ImpulsePhonon}
	\end{figure}
	
	\begin{figure}
    		\centering
    		\includegraphics[width=0.45\textwidth]{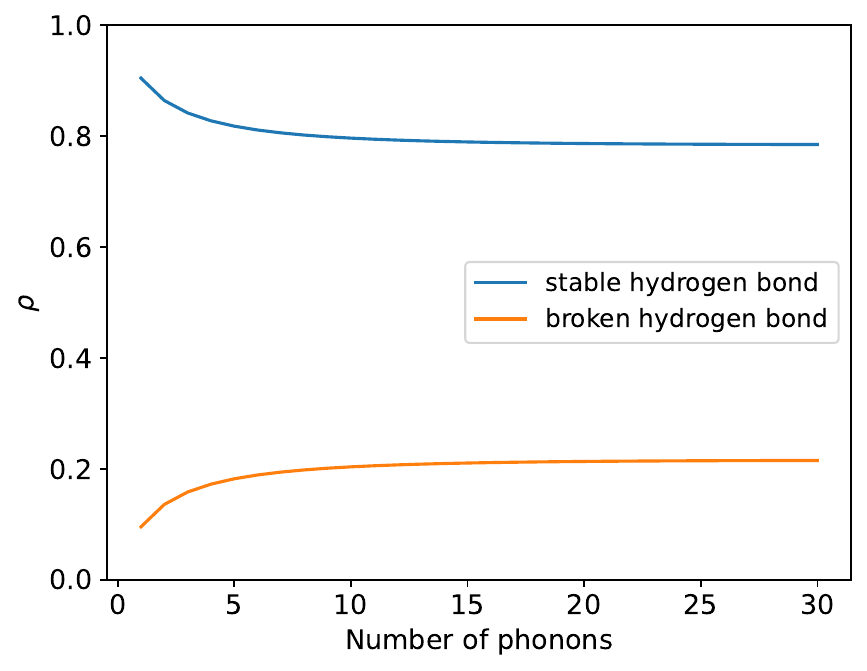}
    		\caption{(online color) {\it Effect of number of phonons on the hydrogen bond formation.} $g_{dist}=g_{prot}=g$, $\gamma_{isol}^{out}=\gamma_{bond}^{out}=\gamma_{ph}^{out}=\gamma$, $\mu_{isol}=\mu_{bond}=0$, $\mu_{phn}=0.01$. The number of phonons is within $[1,\ 30]$.}
    		\label{fig:NumberPhonon}
	\end{figure}

	As mentioned before, the micro-oscillation of protons will affect the stability of hydrogen bonds. Therefore, in this section, the phonon energy to enhance the effect of this micro-oscillation is strengthened, and the changes in the probability of the stable hydrogen bond formation are observed. In order to incorporate the effect of external impulses on proton tunneling into the model, the initial number of phonons is changed (in order to reduce interference factors, the inflow intensities $\gamma_{isol}^{in}$ and $\gamma_{bond}^{in}$ are not considered at this time).
	
	In this subsection, the difference from the previous one is that the initial state is defined as $|0\rangle|0\rangle|N_{phn}\rangle$, where $N_{phn}$ is the number of phonons. Therefore, the evolution involves more than 7 quantum states. In Fig. \ref{fig:ImpulsePhonon}, the result is found that when the number of phonons is large ($N_{phn}\geq 10$), the phonon number changing (external momentum) in the initial state $|0\rangle|0\rangle|N_{phn}\rangle$ has a slight effect on the stability of the hydrogen bond. Even when the number of phonons increases to 30, the probability of the stable hydrogen bond in this model remains stable at around $0.8$. In Fig. \ref{fig:NumberPhonon}, the stability of the hydrogen bond decreases with the increase of the number of phonons. When the number of phonons increases, the probability of the stable hydrogen bond decreases, while the probability of the broken hydrogen bond increases. When the number of phonons is within $[1,\ 10]$, the probabilities of the stable hydrogen bond and the broken hydrogen bond change obviously, but when the phonon number is larger than 10, they change slightly and tend to their respective limits. This is because when the number of a specific type of phonon is large enough, the expected length of the hydrogen bond tends to a stable value; thus, the stability of hydrogen bonds changes slightly. This rule not only appears in the O--H$\cdots$O system but also in the N--H$\cdots$O system \cite{Fontaine2006}.

\begin{figure}
        \centering
        \includegraphics[width=1.\textwidth]{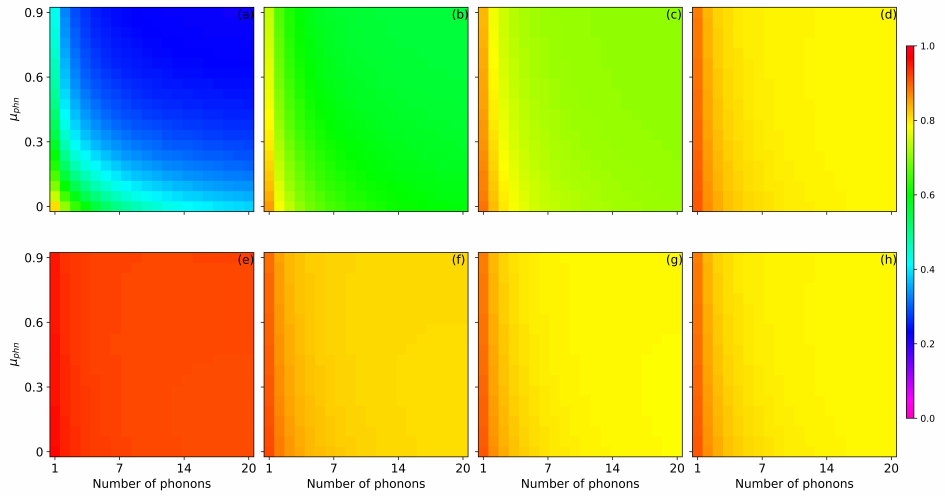}
        \caption{(online color) {\it Effect of the initial phonon number and phonon inflow intensity on the hydrogen bond formation.} $g_{dist}=g_{prot}=g$, $\mu_{isol}=\mu_{bond}=0$, and the value of $\mu_{phn}$ ranges from 0 to 0.9. The initial state is $|0\rangle|0\rangle|N_{phn}\rangle$, $N_{phn}\in[1,\ 20]$. In the first row, $\gamma_{isol}^{out}=\gamma_{phn}^{out}=\gamma$ is fixed, while $\gamma_{bond}^{out}$ is varied in the four panels: $\gamma_{bond}^{out}=0.1\gamma$ in panel (a), $\gamma_{bond}^{out}=0.4\gamma$ in panel (b), $\gamma_{bond}^{out}=0.7\gamma$ in panel (c), and $\gamma_{bond}^{out}=\gamma$ in panel (d). In the second row, $\gamma_{bond}^{out}=\gamma_{phn}^{out}=\gamma$ is fixed, while $\gamma_{isol}^{out}$ is varied in the four panels: $\gamma_{isol}^{out}=0.1\gamma$ in panel (e), $\gamma_{isol}^{out}=0.4\gamma$ in panel (f), $\gamma_{isol}^{out}=0.7\gamma$ in panel (g), and $\gamma_{isol}^{out}=\gamma$ in panel (h).}
        \label{fig:mu+num}
    \end{figure}

	Now, consideration is given to the effect of both the initial phonon number and the phonon inflow intensity on the evolution. In Fig. \ref{fig:mu+num}, the effect of external impulse and phononic temperature (replaced by $\mu_{phn}$) on the probability of the stable hydrogen bond formation is obtained. From the horizontal axis, as the external impulse (the number of phonons in the initial state) increases, the hydrogen bond stability decreases. From the vertical axis, the same conclusion applies to the phononic inflow intensity $\mu_{phn}$. The increase in temperature will make the micro-oscillations of the proton more intense, thereby reducing the stability of hydrogen bonds. From the four panels of the first row in Fig. \ref{fig:mu+num}, as $\gamma_{bond}^{out}$ increases, the probability of the points in the heat map increases, indicating that $\gamma_{bond}^{out}$ promotes the formation of the stable hydrogen bond. In addition, the larger $\gamma_{bond}^{out}$ is, the less obvious the negative effect of $\mu_{phn}$ on the stable hydrogen bond formation is. From the four panels of the second row in Fig. \ref{fig:mu+num}, as $\gamma_{isol}^{out}$ increases, the probability of the points in the heat map decreases, indicating that $\gamma_{bond}^{out}$ hinders the formation of the stable hydrogen bond. In panel (e), $\gamma_{isol}^{out}$ is small enough that the probability of the points in the heat map hardly changes with the increase of the number of phonons; in panels (f)$\sim$(h), the values of $\gamma_{isol}^{out}$ are large, and the probability of the points in each heat map changes with the increase of the number of phonons.
	
    \section{Concluding discussion and future work}
    \label{sec:Conclusion}

    In this paper, the various dynamic aspects of the $\rm{H}_2\rm{O}$-related hydrogen bond model are investigated. And the results show the nontrivial dependence of the hydrogen bond formation on the parameters. Some analytical results of it are derived as follows
	\begin{itemize}
		\item In Sec. \ref{subsec:Comparison}, the periodical oscillations of the closed system are obtained. Once the dissipation of the non-ideal system is introduced, the oscillations disappear, and two final results are obtained: the formation of a stable hydrogen bond or the complete breaking of a hydrogen bond. Different initial states lead to different results: the system tends to form a stable hydrogen bond when the initial state is $|0\rangle|0\rangle|1\rangle$; the system tends to break the hydrogen bond when the initial state is $|1\rangle|1\rangle|0\rangle$.
		\item In Sec. \ref{subsec:EffectInter}, the effect of interactions on evolution is studied. When the initial state is $|0\rangle|0\rangle|1\rangle$, the larger the interaction strengths, the harder it is to form a hydrogen bond, indicating that the interactions hinder the formation of the stable hydrogen bond. On the contrary, when the initial state is $|1\rangle|1\rangle|0\rangle$, the larger the interaction strengths, the easier it is to form a hydrogen bond, indicating that the interactions promote the formation of the stable hydrogen bond. The addition of weak inflows promotes the formation of the hydrogen bond, regardless of whether the initial state is $|1\rangle|1\rangle|0\rangle$ or $|0\rangle|0\rangle|1\rangle$.
		\item In Sec. \ref{subsec:EffectDissi}, the effect of dissipation on the hydrogen bond formation is studied. When the initial state is $|0\rangle|0\rangle|1\rangle$, the probability of the stable hydrogen bond formation is negatively correlated with $\gamma_{isol}^{out}$ and positively correlated with $\gamma_{bond}^{out}$; however, when the initial state is $|1\rangle|1\rangle|0\rangle$, the probability of the stable hydrogen bond formation is negatively correlated with both $\gamma_{isol}^{out}$ and $\gamma_{bond}^{out}$. $\gamma_{phn}^{out}$ promotes the hydrogen bond formation when the initial state is $|0\rangle|0\rangle|1\rangle$, and hinders it when the initial state is $|1\rangle|1\rangle|0\rangle$.
		\item In Sec. \ref{subsec:EffectInflu}, the effect of inflows on the hydrogen bond formation is studied. Regardless of whether the initial state is $|0\rangle|0\rangle|1\rangle$ or $|1\rangle|1\rangle|0\rangle$, the probability of the stable hydrogen bond formation is positively correlated with $\mu_{isol}$ and negatively correlated with $\mu_{bond}$. $\mu_{phn}$ hinders the formation of the hydrogen bond no matter the initial state is $|0\rangle|0\rangle|1\rangle$ or $|1\rangle|1\rangle|0\rangle$.
		\item In Sec. \ref{subsec:external_impulses}, the effect of external impulses on the hydrogen bond formation is obtained. It is evident that the more free phonons flow into the system from the outside, the lower the probability of the stable hydrogen bond formation, and as the number of phonons increases, the probability approaches a limit. The same conclusion as in the previous subsection is obtained: $\mu_{phn}$ hinders the formation of the hydrogen bond in the case of a large number of phonons.
		\item Comparing Secs. \ref{subsec:Comparison} $\sim$ \ref{subsec:external_impulses}, it is obvious that the effect of temperature on quantum evolution and the hydrogen bond formation is far greater than that of other factors.
	\end{itemize}
 
	Based on the above conclusions, our work demonstrates that the evolution and the hydrogen bond formation can be controlled by selectively choosing system parameters. In particular, they are greatly affected by temperature. Although we only studied the dynamic aspects of the simple hydrogen bond model, the results we found will be used as a basis to extend the research to more complex chemical and biological models in the future.

    \section*{CRediT authorship contribution statement}
    J.C. You, R. Chen, and W.S. Li contribute equally to this paper: Investigation (equal), Software (equal), Data curation (equal), Visualization (equal), Formal analysis (equal), Writing --- original draft (supporting); H.-H. Miao: Conceptualization (supporting), Methodology (supporting), Validation (lead), Writing --- original draft (lead), Writing --- review \& editing (equal), Project administration (lead), Supervision (supporting); Y.I. Ozhigov: Conceptualization (lead), Methodology (lead), Resources (lead), Writing --- review \& editing (equal), Supervision (lead).

    \begin{acknowledgments}
	The reported study was funded by the China Scholarship Council, project numbers 202308091509, 202308091210, 202108090327, and 202108090483. The authors acknowledge the Center for Collective Use of Ultra-high-performance Computing Resources (https://parallel.ru/) at Lomonosov Moscow State University for providing supercomputer resources that contributed to the research results presented in this paper.
    \end{acknowledgments}

\appendix

	\section{Quantum master equation for the target model}
    \label{appx:QME}
    
    According to the previous sections, the QME has the following form
    \begin{equation}
         i\hbar\dot{\rho} =  [H, \rho] + i\mathcal{L}(\rho),
         \label{appxeq:QME}
    \end{equation}
    where $H$ is the Hamiltonian, $\rho$ is the density matrix, and $\mathcal{L}=L^{in} + L^{out}$. Then
    	\begin{equation}
    		\begin{aligned}
        	&L^{in}(\rho)=\gamma_{bond}^{in}\left( A_{bond}^\dag\rho A_{bond} - \frac{1}{2}\left(  A_{bond} A_{bond}^\dag \rho + \rho A_{bond} A_{bond}^\dag  \right) \right)\\
        	&+\gamma_{isol}^{in}\left( A_{isol}^\dag\rho A_{isol} - \frac{1}{2}\left(  A_{isol} A_{isol}^\dag \rho + \rho A_{isol} A_{isol}^\dag  \right) \right)+\gamma_{phn}^{in}\left( A_{phn}^\dag\rho A_{phn} - \frac{1}{2}\left(  A_{phnl} A_{phn}^\dag \rho + \rho A_{phn} A_{phn}^\dag  \right) \right),
        	\end{aligned}
        	\label{appxeq:Linblad_in}
    	\end{equation}
    	\begin{equation}
    		\begin{aligned}
         &L^{out}(\rho)=\gamma_{bond}^{out}\left( A_{bond}\rho A_{bond}^\dag - \frac{1}{2}\left( A_{bond}^\dag A_{bond}\rho + \rho A_{bond}^\dag A_{bond}  \right) \right)\\
         &+\gamma_{isol}^{out}\left( A_{isol}\rho A_{isol}^\dag - \frac{1}{2}\left( A_{isol}^\dag A_{isol}\rho + \rho A_{isol}^\dag A_{isol}  \right) \right)+\gamma_{phn}^{out}\left( A_{phn}\rho A_{phn}^\dag - \frac{1}{2}\left( A_{phn}^\dag A_{phn}\rho + \rho A_{phn}^\dag A_{phn}  \right) \right),
        	\label{appxeq:Linblad_out}
        	\end{aligned}
    	\end{equation}
    where $\gamma_{bond}^{in}=\mu_{bond}\gamma_{bond}^{out},\ \gamma_{isol}^{in}=\mu_{isol}\gamma_{isol}^{out},\ \gamma_{phn}^{in}=\mu_{phn}\gamma_{phn}^{out}$. Here, $\mu_{bond}<1,\ \mu_{isol}<1$, and $\mu_{phn}<1$, thus, $\gamma_{bond}^{in}<\gamma_{bond}^{out},\ \gamma_{isol}^{in}<\gamma_{isol}^{out}$, and $\gamma_{phn}^{in}<\gamma_{phn}^{out}$.

    \section{Estimation of physical quantities}
    \label{appx:PhysQuantities}
    
    The average energy of hydrogen bonds in water is $E_{bond}=21[\rm{kJ/mol}]$. To convert this energy to the energy of a single hydrogen bond, estimate it by using Avogadro's constant $N_A$ --- the number of particles per mole of substance, which is approximately ($6.022 \times 10^{23}$). Through calculation, it can be estimated that the energy of a single hydrogen bond
    \begin{equation}
    		E_{H-bond}=\frac{E_{bond}}{N_A}=\frac{21\times 10^{3}\ \text{[J/mol}]}{6.022\times 10^{23}\text{[molecules/mol]}}\approx 3.32\times 10^{-20}\ [\rm{J}] = 0.217655\ \text{[eV]}. 
    \end{equation}
    Replacing $E$ in $\tau =\frac{\hbar}{E}$ with $E_{H-bond}$, the characteristic time scale is obtained as follows
	\begin{equation}
    		\tau = \frac{\hbar}{E_{H-bond}} = \frac{\hbar \times N_A}{E_{bond}}= \frac{1.0545718\times 10^{-34}\ [J\cdot s] \times 6.022\times 10^{23}\text{[molecules/mol}]}{21\times 10^{3}\ \text{[J/mol]}} \approx 3.175316 \times 10^{-15}\ [\rm{s}].  
    \end{equation}

	\bibliography{bibliography}

\begin{thebibliography}{61}%
\makeatletter
\providecommand \@ifxundefined [1]{%
 \@ifx{#1\undefined}
}%
\providecommand \@ifnum [1]{%
 \ifnum #1\expandafter \@firstoftwo
 \else \expandafter \@secondoftwo
 \fi
}%
\providecommand \@ifx [1]{%
 \ifx #1\expandafter \@firstoftwo
 \else \expandafter \@secondoftwo
 \fi
}%
\providecommand \natexlab [1]{#1}%
\providecommand \enquote  [1]{``#1''}%
\providecommand \bibnamefont  [1]{#1}%
\providecommand \bibfnamefont [1]{#1}%
\providecommand \citenamefont [1]{#1}%
\providecommand \href@noop [0]{\@secondoftwo}%
\providecommand \href [0]{\begingroup \@sanitize@url \@href}%
\providecommand \@href[1]{\@@startlink{#1}\@@href}%
\providecommand \@@href[1]{\endgroup#1\@@endlink}%
\providecommand \@sanitize@url [0]{\catcode `\\12\catcode `\$12\catcode
  `\&12\catcode `\#12\catcode `\^12\catcode `\_12\catcode `\%12\relax}%
\providecommand \@@startlink[1]{}%
\providecommand \@@endlink[0]{}%
\providecommand \url  [0]{\begingroup\@sanitize@url \@url }%
\providecommand \@url [1]{\endgroup\@href {#1}{\urlprefix }}%
\providecommand \urlprefix  [0]{URL }%
\providecommand \Eprint [0]{\href }%
\providecommand \doibase [0]{https://doi.org/}%
\providecommand \selectlanguage [0]{\@gobble}%
\providecommand \bibinfo  [0]{\@secondoftwo}%
\providecommand \bibfield  [0]{\@secondoftwo}%
\providecommand \translation [1]{[#1]}%
\providecommand \BibitemOpen [0]{}%
\providecommand \bibitemStop [0]{}%
\providecommand \bibitemNoStop [0]{.\EOS\space}%
\providecommand \EOS [0]{\spacefactor3000\relax}%
\providecommand \BibitemShut  [1]{\csname bibitem#1\endcsname}%
\let\auto@bib@innerbib\@empty
\bibitem [{\citenamefont {McArdle}\ \emph {et~al.}(2020)\citenamefont
  {McArdle}, \citenamefont {Endo}, \citenamefont {Aspuru-Guzik}, \citenamefont
  {Benjamin},\ and\ \citenamefont {Yuan}}]{McArdle2020}%
  \BibitemOpen
  \bibfield  {author} {\bibinfo {author} {\bibfnamefont {S.}~\bibnamefont
  {McArdle}}, \bibinfo {author} {\bibfnamefont {S.}~\bibnamefont {Endo}},
  \bibinfo {author} {\bibfnamefont {A.}~\bibnamefont {Aspuru-Guzik}}, \bibinfo
  {author} {\bibfnamefont {S.~C.}\ \bibnamefont {Benjamin}},\ and\ \bibinfo
  {author} {\bibfnamefont {X.}~\bibnamefont {Yuan}},\ }\bibfield  {title}
  {\bibinfo {title} {Quantum computational chemistry},\ }\href
  {https://doi.org/10.1103/RevModPhys.92.015003} {\bibfield  {journal}
  {\bibinfo  {journal} {Rev. Mod. Phys.}\ }\textbf {\bibinfo {volume} {92}},\
  \bibinfo {pages} {015003} (\bibinfo {year} {2020})}\BibitemShut {NoStop}%
\bibitem [{\citenamefont {Baiardi}\ \emph {et~al.}(2023)\citenamefont
  {Baiardi}, \citenamefont {Christandl},\ and\ \citenamefont
  {Reiher}}]{Baiardi2023}%
  \BibitemOpen
  \bibfield  {author} {\bibinfo {author} {\bibfnamefont {A.}~\bibnamefont
  {Baiardi}}, \bibinfo {author} {\bibfnamefont {M.}~\bibnamefont
  {Christandl}},\ and\ \bibinfo {author} {\bibfnamefont {M.}~\bibnamefont
  {Reiher}},\ }\bibfield  {title} {\bibinfo {title} {Quantum computing for
  molecular biology},\ }\href
  {https://doi.org/https://doi.org/10.1002/cbic.202300120} {\bibfield
  {journal} {\bibinfo  {journal} {ChemBioChem}\ }\textbf {\bibinfo {volume}
  {24}},\ \bibinfo {pages} {e202300120} (\bibinfo {year} {2023})}\BibitemShut
  {NoStop}%
\bibitem [{\citenamefont {Albuquerque}\ \emph {et~al.}(2021)\citenamefont
  {Albuquerque}, \citenamefont {Fulco}, \citenamefont {Caetano},\ and\
  \citenamefont {Freire}}]{Albuquerque2021}%
  \BibitemOpen
  \bibfield  {author} {\bibinfo {author} {\bibfnamefont {E.}~\bibnamefont
  {Albuquerque}}, \bibinfo {author} {\bibfnamefont {U.}~\bibnamefont {Fulco}},
  \bibinfo {author} {\bibfnamefont {E.}~\bibnamefont {Caetano}},\ and\ \bibinfo
  {author} {\bibfnamefont {V.}~\bibnamefont {Freire}},\ }\href@noop {} {\emph
  {\bibinfo {title} {Quantum Chemistry Simulation of Biological Molecules}}}\
  (\bibinfo  {publisher} {Cambridge University Press},\ \bibinfo {year}
  {2021})\BibitemShut {NoStop}%
\bibitem [{\citenamefont {Bellman}(1957)}]{Bellman1957}%
  \BibitemOpen
  \bibfield  {author} {\bibinfo {author} {\bibfnamefont {R.~E.}\ \bibnamefont
  {Bellman}},\ }\href@noop {} {\emph {\bibinfo {title} {Dynamic Programming}}}\
  (\bibinfo  {publisher} {Princeton University Press},\ \bibinfo {year}
  {1957})\BibitemShut {NoStop}%
\bibitem [{\citenamefont {Bellman}(1961)}]{Bellman1961}%
  \BibitemOpen
  \bibfield  {author} {\bibinfo {author} {\bibfnamefont {R.~E.}\ \bibnamefont
  {Bellman}},\ }\href@noop {} {\emph {\bibinfo {title} {Adaptive control
  processes: a guided tour}}}\ (\bibinfo  {publisher} {Princeton University
  Press},\ \bibinfo {year} {1961})\BibitemShut {NoStop}%
\bibitem [{\citenamefont {Moore}\ and\ \citenamefont
  {Winmill}(1912)}]{Moore1912}%
  \BibitemOpen
  \bibfield  {author} {\bibinfo {author} {\bibfnamefont {T.~S.}\ \bibnamefont
  {Moore}}\ and\ \bibinfo {author} {\bibfnamefont {T.~F.}\ \bibnamefont
  {Winmill}},\ }\bibfield  {title} {\bibinfo {title} {{CLXXVII}.—the state of
  amines in aqueous solution},\ }\href {https://doi.org/10.1039/CT9120101635}
  {\bibfield  {journal} {\bibinfo  {journal} {J. Chem. Soc.{,} Trans.}\
  }\textbf {\bibinfo {volume} {101}},\ \bibinfo {pages} {1635} (\bibinfo {year}
  {1912})}\BibitemShut {NoStop}%
\bibitem [{\citenamefont {Latimer}\ and\ \citenamefont
  {Rodebush}(1920)}]{Latimer1920}%
  \BibitemOpen
  \bibfield  {author} {\bibinfo {author} {\bibfnamefont {W.~M.}\ \bibnamefont
  {Latimer}}\ and\ \bibinfo {author} {\bibfnamefont {W.~H.}\ \bibnamefont
  {Rodebush}},\ }\bibfield  {title} {\bibinfo {title} {Polarity and ionization
  from the standpoint of the lewis theory of valence.},\ }\href
  {https://doi.org/10.1021/ja01452a015} {\bibfield  {journal} {\bibinfo
  {journal} {Journal of the American Chemical Society}\ }\textbf {\bibinfo
  {volume} {42}},\ \bibinfo {pages} {1419} (\bibinfo {year}
  {1920})}\BibitemShut {NoStop}%
\bibitem [{\citenamefont {Dillon}(2012)}]{Dillon2012}%
  \BibitemOpen
  \bibfield  {author} {\bibinfo {author} {\bibfnamefont {P.~F.}\ \bibnamefont
  {Dillon}},\ }\href
  {https://www.cambridge.org/ru/universitypress/subjects/life-sciences/biophysics-and-physiology/biophysics-physiological-approach?format=PB&isbn=9780521172165}
  {\emph {\bibinfo {title} {Biophysics: A Physiological Approach}}}\ (\bibinfo
  {publisher} {Cambridge University Press},\ \bibinfo {year}
  {2012})\BibitemShut {NoStop}%
\bibitem [{\citenamefont {Stahl}\ and\ \citenamefont
  {Jencks}(1986)}]{Stahl1986}%
  \BibitemOpen
  \bibfield  {author} {\bibinfo {author} {\bibfnamefont {N.}~\bibnamefont
  {Stahl}}\ and\ \bibinfo {author} {\bibfnamefont {W.~P.}\ \bibnamefont
  {Jencks}},\ }\bibfield  {title} {\bibinfo {title} {Hydrogen bonding between
  solutes in aqueous solution},\ }\href {https://doi.org/10.1021/ja00274a058}
  {\bibfield  {journal} {\bibinfo  {journal} {Journal of the American Chemical
  Society}\ }\textbf {\bibinfo {volume} {108}},\ \bibinfo {pages} {4196}
  (\bibinfo {year} {1986})}\BibitemShut {NoStop}%
\bibitem [{\citenamefont {Gustin}\ \emph {et~al.}(2023)\citenamefont {Gustin},
  \citenamefont {Kim}, \citenamefont {McCamant},\ and\ \citenamefont
  {Franco}}]{Ignacio2023}%
  \BibitemOpen
  \bibfield  {author} {\bibinfo {author} {\bibfnamefont {I.}~\bibnamefont
  {Gustin}}, \bibinfo {author} {\bibfnamefont {C.~W.}\ \bibnamefont {Kim}},
  \bibinfo {author} {\bibfnamefont {D.~W.}\ \bibnamefont {McCamant}},\ and\
  \bibinfo {author} {\bibfnamefont {I.}~\bibnamefont {Franco}},\ }\bibfield
  {title} {\bibinfo {title} {Mapping electronic decoherence pathways in
  molecules},\ }\href {https://doi.org/10.1073/pnas.2309987120} {\bibfield
  {journal} {\bibinfo  {journal} {Proceedings of the National Academy of
  Sciences}\ }\textbf {\bibinfo {volume} {120}},\ \bibinfo {pages}
  {e2309987120} (\bibinfo {year} {2023})}\BibitemShut {NoStop}%
\bibitem [{\citenamefont {He}\ \emph {et~al.}(2022)\citenamefont {He},
  \citenamefont {Li}, \citenamefont {Castelli}, \citenamefont {Li},
  \citenamefont {Zhang}, \citenamefont {Zhang}, \citenamefont {Li},
  \citenamefont {Wang}, \citenamefont {Gao}, \citenamefont {Peng},
  \citenamefont {Hou}, \citenamefont {Shen}, \citenamefont {L\"u},
  \citenamefont {Wu}, \citenamefont {Hedeg\aa{}rd},\ and\ \citenamefont
  {Wang}}]{He2022}%
  \BibitemOpen
  \bibfield  {author} {\bibinfo {author} {\bibfnamefont {Y.}~\bibnamefont
  {He}}, \bibinfo {author} {\bibfnamefont {N.}~\bibnamefont {Li}}, \bibinfo
  {author} {\bibfnamefont {I.~E.}\ \bibnamefont {Castelli}}, \bibinfo {author}
  {\bibfnamefont {R.}~\bibnamefont {Li}}, \bibinfo {author} {\bibfnamefont
  {Y.}~\bibnamefont {Zhang}}, \bibinfo {author} {\bibfnamefont
  {X.}~\bibnamefont {Zhang}}, \bibinfo {author} {\bibfnamefont
  {C.}~\bibnamefont {Li}}, \bibinfo {author} {\bibfnamefont {B.}~\bibnamefont
  {Wang}}, \bibinfo {author} {\bibfnamefont {S.}~\bibnamefont {Gao}}, \bibinfo
  {author} {\bibfnamefont {L.}~\bibnamefont {Peng}}, \bibinfo {author}
  {\bibfnamefont {S.}~\bibnamefont {Hou}}, \bibinfo {author} {\bibfnamefont
  {Z.}~\bibnamefont {Shen}}, \bibinfo {author} {\bibfnamefont {J.-T.}\
  \bibnamefont {L\"u}}, \bibinfo {author} {\bibfnamefont {K.}~\bibnamefont
  {Wu}}, \bibinfo {author} {\bibfnamefont {P.}~\bibnamefont {Hedeg\aa{}rd}},\
  and\ \bibinfo {author} {\bibfnamefont {Y.}~\bibnamefont {Wang}},\ }\bibfield
  {title} {\bibinfo {title} {Observation of biradical spin coupling through
  hydrogen bonds},\ }\href {https://doi.org/10.1103/PhysRevLett.128.236401}
  {\bibfield  {journal} {\bibinfo  {journal} {Phys. Rev. Lett.}\ }\textbf
  {\bibinfo {volume} {128}},\ \bibinfo {pages} {236401} (\bibinfo {year}
  {2022})}\BibitemShut {NoStop}%
\bibitem [{\citenamefont {Georgiev}\ and\ \citenamefont
  {Glazebrook}(2022)}]{Danko2022}%
  \BibitemOpen
  \bibfield  {author} {\bibinfo {author} {\bibfnamefont {D.~D.}\ \bibnamefont
  {Georgiev}}\ and\ \bibinfo {author} {\bibfnamefont {J.~F.}\ \bibnamefont
  {Glazebrook}},\ }\bibfield  {title} {\bibinfo {title} {Thermal stability of
  solitons in protein $\alpha$-helices},\ }\href
  {https://doi.org/https://doi.org/10.1016/j.chaos.2021.111644} {\bibfield
  {journal} {\bibinfo  {journal} {Chaos, Solitons \& Fractals}\ }\textbf
  {\bibinfo {volume} {155}},\ \bibinfo {pages} {111644} (\bibinfo {year}
  {2022})}\BibitemShut {NoStop}%
\bibitem [{\citenamefont {Di~Liberto}\ \emph {et~al.}(2018)\citenamefont
  {Di~Liberto}, \citenamefont {Conte},\ and\ \citenamefont {Ceotto}}]{Di2018}%
  \BibitemOpen
  \bibfield  {author} {\bibinfo {author} {\bibfnamefont {G.}~\bibnamefont
  {Di~Liberto}}, \bibinfo {author} {\bibfnamefont {R.}~\bibnamefont {Conte}},\
  and\ \bibinfo {author} {\bibfnamefont {M.}~\bibnamefont {Ceotto}},\
  }\bibfield  {title} {\bibinfo {title} {{“Divide-and-conquer”
  semiclassical molecular dynamics: An application to water clusters}},\ }\href
  {https://doi.org/10.1063/1.5023155} {\bibfield  {journal} {\bibinfo
  {journal} {The Journal of Chemical Physics}\ }\textbf {\bibinfo {volume}
  {148}},\ \bibinfo {pages} {104302} (\bibinfo {year} {2018})}\BibitemShut
  {NoStop}%
\bibitem [{\citenamefont {Yamada}\ and\ \citenamefont
  {Tada}(2020)}]{Yamada2020}%
  \BibitemOpen
  \bibfield  {author} {\bibinfo {author} {\bibfnamefont {M.~G.}\ \bibnamefont
  {Yamada}}\ and\ \bibinfo {author} {\bibfnamefont {Y.}~\bibnamefont {Tada}},\
  }\bibfield  {title} {\bibinfo {title} {Quantum valence bond ice theory for
  proton-driven quantum spin-dipole liquids},\ }\href
  {https://doi.org/10.1103/PhysRevResearch.2.043077} {\bibfield  {journal}
  {\bibinfo  {journal} {Phys. Rev. Res.}\ }\textbf {\bibinfo {volume} {2}},\
  \bibinfo {pages} {043077} (\bibinfo {year} {2020})}\BibitemShut {NoStop}%
\bibitem [{\citenamefont {Pusuluk}\ \emph
  {et~al.}(2018{\natexlab{a}})\citenamefont {Pusuluk}, \citenamefont {Torun},\
  and\ \citenamefont {Deliduman}}]{Pusuluk2018}%
  \BibitemOpen
  \bibfield  {author} {\bibinfo {author} {\bibfnamefont {O.}~\bibnamefont
  {Pusuluk}}, \bibinfo {author} {\bibfnamefont {G.}~\bibnamefont {Torun}},\
  and\ \bibinfo {author} {\bibfnamefont {C.}~\bibnamefont {Deliduman}},\
  }\bibfield  {title} {\bibinfo {title} {Quantum entanglement shared in
  hydrogen bonds and its usage as a resource in molecular recognition},\ }\href
  {https://doi.org/10.1142/S0217984918503086} {\bibfield  {journal} {\bibinfo
  {journal} {Modern Physics Letters B}\ }\textbf {\bibinfo {volume} {32}},\
  \bibinfo {pages} {1850308} (\bibinfo {year}
  {2018}{\natexlab{a}})}\BibitemShut {NoStop}%
\bibitem [{\citenamefont {Pusuluk}\ \emph
  {et~al.}(2018{\natexlab{b}})\citenamefont {Pusuluk}, \citenamefont {Farrow},
  \citenamefont {Deliduman}, \citenamefont {Burnett},\ and\ \citenamefont
  {Vedral}}]{Farrow2018}%
  \BibitemOpen
  \bibfield  {author} {\bibinfo {author} {\bibfnamefont {O.}~\bibnamefont
  {Pusuluk}}, \bibinfo {author} {\bibfnamefont {T.}~\bibnamefont {Farrow}},
  \bibinfo {author} {\bibfnamefont {C.}~\bibnamefont {Deliduman}}, \bibinfo
  {author} {\bibfnamefont {K.}~\bibnamefont {Burnett}},\ and\ \bibinfo {author}
  {\bibfnamefont {V.}~\bibnamefont {Vedral}},\ }\bibfield  {title} {\bibinfo
  {title} {Proton tunnelling in hydrogen bonds and its implications in an
  induced-fit model of enzyme catalysis},\ }\href
  {https://doi.org/10.1098/rspa.2018.0037} {\bibfield  {journal} {\bibinfo
  {journal} {Proceedings of the Royal Society A: Mathematical, Physical and
  Engineering Sciences}\ }\textbf {\bibinfo {volume} {474}},\ \bibinfo {pages}
  {20180037} (\bibinfo {year} {2018}{\natexlab{b}})}\BibitemShut {NoStop}%
\bibitem [{\citenamefont {Rabi}(1936)}]{Rabi1936}%
  \BibitemOpen
  \bibfield  {author} {\bibinfo {author} {\bibfnamefont {I.~I.}\ \bibnamefont
  {Rabi}},\ }\bibfield  {title} {\bibinfo {title} {On the process of space
  quantization},\ }\href {https://doi.org/10.1103/PhysRev.49.324} {\bibfield
  {journal} {\bibinfo  {journal} {Phys. Rev.}\ }\textbf {\bibinfo {volume}
  {49}},\ \bibinfo {pages} {324} (\bibinfo {year} {1936})}\BibitemShut
  {NoStop}%
\bibitem [{\citenamefont {Rabi}(1937)}]{Rabi1937}%
  \BibitemOpen
  \bibfield  {author} {\bibinfo {author} {\bibfnamefont {I.~I.}\ \bibnamefont
  {Rabi}},\ }\bibfield  {title} {\bibinfo {title} {Space quantization in a
  gyrating magnetic field},\ }\href {https://doi.org/10.1103/PhysRev.51.652}
  {\bibfield  {journal} {\bibinfo  {journal} {Phys. Rev.}\ }\textbf {\bibinfo
  {volume} {51}},\ \bibinfo {pages} {652} (\bibinfo {year} {1937})}\BibitemShut
  {NoStop}%
\bibitem [{\citenamefont {Dicke}(1954)}]{Dicke1954}%
  \BibitemOpen
  \bibfield  {author} {\bibinfo {author} {\bibfnamefont {R.~H.}\ \bibnamefont
  {Dicke}},\ }\bibfield  {title} {\bibinfo {title} {Coherence in spontaneous
  radiation processes},\ }\href {https://doi.org/10.1103/PhysRev.93.99}
  {\bibfield  {journal} {\bibinfo  {journal} {Phys. Rev.}\ }\textbf {\bibinfo
  {volume} {93}},\ \bibinfo {pages} {99} (\bibinfo {year} {1954})}\BibitemShut
  {NoStop}%
\bibitem [{\citenamefont {Hopfield}(1958)}]{Hopfield1958}%
  \BibitemOpen
  \bibfield  {author} {\bibinfo {author} {\bibfnamefont {J.~J.}\ \bibnamefont
  {Hopfield}},\ }\bibfield  {title} {\bibinfo {title} {Theory of the
  contribution of excitons to the complex dielectric constant of crystals},\
  }\href {https://doi.org/10.1103/PhysRev.112.1555} {\bibfield  {journal}
  {\bibinfo  {journal} {Phys. Rev.}\ }\textbf {\bibinfo {volume} {112}},\
  \bibinfo {pages} {1555} (\bibinfo {year} {1958})}\BibitemShut {NoStop}%
\bibitem [{\citenamefont {Casanova}\ \emph {et~al.}(2010)\citenamefont
  {Casanova}, \citenamefont {Romero}, \citenamefont {Lizuain}, \citenamefont
  {Garc\'{\i}a-Ripoll},\ and\ \citenamefont {Solano}}]{Casanova2010}%
  \BibitemOpen
  \bibfield  {author} {\bibinfo {author} {\bibfnamefont {J.}~\bibnamefont
  {Casanova}}, \bibinfo {author} {\bibfnamefont {G.}~\bibnamefont {Romero}},
  \bibinfo {author} {\bibfnamefont {I.}~\bibnamefont {Lizuain}}, \bibinfo
  {author} {\bibfnamefont {J.~J.}\ \bibnamefont {Garc\'{\i}a-Ripoll}},\ and\
  \bibinfo {author} {\bibfnamefont {E.}~\bibnamefont {Solano}},\ }\bibfield
  {title} {\bibinfo {title} {Deep strong coupling regime of the jaynes-cummings
  model},\ }\href {https://doi.org/10.1103/PhysRevLett.105.263603} {\bibfield
  {journal} {\bibinfo  {journal} {Phys. Rev. Lett.}\ }\textbf {\bibinfo
  {volume} {105}},\ \bibinfo {pages} {263603} (\bibinfo {year}
  {2010})}\BibitemShut {NoStop}%
\bibitem [{\citenamefont {Jaynes}\ and\ \citenamefont
  {Cummings}(1963)}]{Jaynes1963}%
  \BibitemOpen
  \bibfield  {author} {\bibinfo {author} {\bibfnamefont {E.~T.}\ \bibnamefont
  {Jaynes}}\ and\ \bibinfo {author} {\bibfnamefont {F.~W.}\ \bibnamefont
  {Cummings}},\ }\bibfield  {title} {\bibinfo {title} {Comparison of quantum
  and semiclassical radiation theories with application to the beam maser},\
  }\href {https://doi.org/10.1109/PROC.1963.1664} {\bibfield  {journal}
  {\bibinfo  {journal} {Proceedings of the IEEE}\ }\textbf {\bibinfo {volume}
  {51}},\ \bibinfo {pages} {89} (\bibinfo {year} {1963})}\BibitemShut {NoStop}%
\bibitem [{\citenamefont {Tavis}\ and\ \citenamefont
  {Cummings}(1968)}]{Tavis1968}%
  \BibitemOpen
  \bibfield  {author} {\bibinfo {author} {\bibfnamefont {M.}~\bibnamefont
  {Tavis}}\ and\ \bibinfo {author} {\bibfnamefont {F.~W.}\ \bibnamefont
  {Cummings}},\ }\bibfield  {title} {\bibinfo {title} {Exact solution for an
  $n$-molecule---radiation-field {H}amiltonian},\ }\href
  {https://doi.org/10.1103/PhysRev.170.379} {\bibfield  {journal} {\bibinfo
  {journal} {Phys. Rev.}\ }\textbf {\bibinfo {volume} {170}},\ \bibinfo {pages}
  {379} (\bibinfo {year} {1968})}\BibitemShut {NoStop}%
\bibitem [{\citenamefont {Angelakis}\ \emph {et~al.}(2007)\citenamefont
  {Angelakis}, \citenamefont {Santos},\ and\ \citenamefont
  {Bose}}]{Angelakis2007}%
  \BibitemOpen
  \bibfield  {author} {\bibinfo {author} {\bibfnamefont {D.~G.}\ \bibnamefont
  {Angelakis}}, \bibinfo {author} {\bibfnamefont {M.~F.}\ \bibnamefont
  {Santos}},\ and\ \bibinfo {author} {\bibfnamefont {S.}~\bibnamefont {Bose}},\
  }\bibfield  {title} {\bibinfo {title} {Photon-blockade-induced mott
  transitions and {$XY$} spin models in coupled cavity arrays},\ }\href
  {https://doi.org/10.1103/PhysRevA.76.031805} {\bibfield  {journal} {\bibinfo
  {journal} {Phys. Rev. A}\ }\textbf {\bibinfo {volume} {76}},\ \bibinfo
  {pages} {031805} (\bibinfo {year} {2007})}\BibitemShut {NoStop}%
\bibitem [{\citenamefont {Ozhigov}(2020)}]{OzhigovYI2020}%
  \BibitemOpen
  \bibfield  {author} {\bibinfo {author} {\bibfnamefont {Y.}~\bibnamefont
  {Ozhigov}},\ }\bibfield  {title} {\bibinfo {title} {Quantum gates on
  asynchronous atomic excitations},\ }\href {https://doi.org/10.1070/QEL17320}
  {\bibfield  {journal} {\bibinfo  {journal} {Quantum Electronics}\ }\textbf
  {\bibinfo {volume} {50}},\ \bibinfo {pages} {947} (\bibinfo {year}
  {2020})}\BibitemShut {NoStop}%
\bibitem [{\citenamefont {Düll}\ \emph {et~al.}(2021)\citenamefont {Düll},
  \citenamefont {Kulagin}, \citenamefont {Lee}, \citenamefont {Ozhigov},
  \citenamefont {Miao},\ and\ \citenamefont {Zheng}}]{Dull2021}%
  \BibitemOpen
  \bibfield  {author} {\bibinfo {author} {\bibfnamefont {R.}~\bibnamefont
  {Düll}}, \bibinfo {author} {\bibfnamefont {A.}~\bibnamefont {Kulagin}},
  \bibinfo {author} {\bibfnamefont {L.}~\bibnamefont {Lee}}, \bibinfo {author}
  {\bibfnamefont {Y.}~\bibnamefont {Ozhigov}}, \bibinfo {author} {\bibfnamefont
  {H.}~\bibnamefont {Miao}},\ and\ \bibinfo {author} {\bibfnamefont
  {K.}~\bibnamefont {Zheng}},\ }\bibfield  {title} {\bibinfo {title} {Quality
  of control in the {T}avis--{C}ummings--{H}ubbard model},\ }\href
  {https://doi.org/10.1007/s10598-021-09517-y} {\bibfield  {journal} {\bibinfo
  {journal} {Computational Mathematics and Modeling}\ }\textbf {\bibinfo
  {volume} {32}},\ \bibinfo {pages} {75} (\bibinfo {year} {2021})}\BibitemShut
  {NoStop}%
\bibitem [{\citenamefont {Smith}\ \emph {et~al.}(2021)\citenamefont {Smith},
  \citenamefont {Bhattacharya},\ and\ \citenamefont {Masiello}}]{Smith2021}%
  \BibitemOpen
  \bibfield  {author} {\bibinfo {author} {\bibfnamefont {K.~C.}\ \bibnamefont
  {Smith}}, \bibinfo {author} {\bibfnamefont {A.}~\bibnamefont
  {Bhattacharya}},\ and\ \bibinfo {author} {\bibfnamefont {D.~J.}\ \bibnamefont
  {Masiello}},\ }\bibfield  {title} {\bibinfo {title} {Exact $k$-body
  representation of the {J}aynes--{C}ummings interaction in the dressed basis:
  Insight into many-body phenomena with light},\ }\href
  {https://doi.org/10.1103/PhysRevA.104.013707} {\bibfield  {journal} {\bibinfo
   {journal} {Phys. Rev. A}\ }\textbf {\bibinfo {volume} {104}},\ \bibinfo
  {pages} {013707} (\bibinfo {year} {2021})}\BibitemShut {NoStop}%
\bibitem [{\citenamefont {Miao}(2024)}]{MiaoHuihui2024}%
  \BibitemOpen
  \bibfield  {author} {\bibinfo {author} {\bibfnamefont {H.-h.}\ \bibnamefont
  {Miao}},\ }\bibfield  {title} {\bibinfo {title} {Investigating entropic
  dynamics of multiqubit cavity qed system},\ }\href
  {https://doi.org/https://doi.org/10.1002/qute.202400246} {\bibfield
  {journal} {\bibinfo  {journal} {Advanced Quantum Technologies}\ }\textbf
  {\bibinfo {volume} {7}},\ \bibinfo {pages} {2400246} (\bibinfo {year}
  {2024})}\BibitemShut {NoStop}%
\bibitem [{\citenamefont {Miao}\ and\ \citenamefont {Li}(2025)}]{MiaoLi2025}%
  \BibitemOpen
  \bibfield  {author} {\bibinfo {author} {\bibfnamefont {H.-h.}\ \bibnamefont
  {Miao}}\ and\ \bibinfo {author} {\bibfnamefont {W.}~\bibnamefont {Li}},\
  }\bibfield  {title} {\bibinfo {title} {Entanglement and quantum discord in
  the cavity {QED} model},\ }\href
  {https://doi.org/10.1016/j.heliyon.2024.e41194} {\bibfield  {journal}
  {\bibinfo  {journal} {Heliyon}\ }\textbf {\bibinfo {volume} {11}},\ \bibinfo
  {pages} {e41194} (\bibinfo {year} {2025})}\BibitemShut {NoStop}%
\bibitem [{\citenamefont {Lee}\ \emph {et~al.}(1999)\citenamefont {Lee},
  \citenamefont {Geckeler}, \citenamefont {Heurich}, \citenamefont {Gupta},
  \citenamefont {Cheong}, \citenamefont {Secrest},\ and\ \citenamefont
  {Meystre}}]{Lee1999}%
  \BibitemOpen
  \bibfield  {author} {\bibinfo {author} {\bibfnamefont {E.~S.}\ \bibnamefont
  {Lee}}, \bibinfo {author} {\bibfnamefont {C.}~\bibnamefont {Geckeler}},
  \bibinfo {author} {\bibfnamefont {J.}~\bibnamefont {Heurich}}, \bibinfo
  {author} {\bibfnamefont {A.}~\bibnamefont {Gupta}}, \bibinfo {author}
  {\bibfnamefont {K.-I.}\ \bibnamefont {Cheong}}, \bibinfo {author}
  {\bibfnamefont {S.}~\bibnamefont {Secrest}},\ and\ \bibinfo {author}
  {\bibfnamefont {P.}~\bibnamefont {Meystre}},\ }\bibfield  {title} {\bibinfo
  {title} {Dark states of dressed bose-einstein condensates},\ }\href
  {https://doi.org/10.1103/PhysRevA.60.4006} {\bibfield  {journal} {\bibinfo
  {journal} {Phys. Rev. A}\ }\textbf {\bibinfo {volume} {60}},\ \bibinfo
  {pages} {4006} (\bibinfo {year} {1999})}\BibitemShut {NoStop}%
\bibitem [{\citenamefont {André}\ \emph {et~al.}(2002)\citenamefont {André},
  \citenamefont {Duan},\ and\ \citenamefont {Lukin}}]{Andre2002}%
  \BibitemOpen
  \bibfield  {author} {\bibinfo {author} {\bibfnamefont {A.}~\bibnamefont
  {André}}, \bibinfo {author} {\bibfnamefont {L.-M.}\ \bibnamefont {Duan}},\
  and\ \bibinfo {author} {\bibfnamefont {M.~D.}\ \bibnamefont {Lukin}},\
  }\bibfield  {title} {\bibinfo {title} {Coherent atom interactions mediated by
  dark-state polaritons},\ }\href
  {https://doi.org/10.1103/PhysRevLett.88.243602} {\bibfield  {journal}
  {\bibinfo  {journal} {Phys Rev Lett}\ }\textbf {\bibinfo {volume} {88}},\
  \bibinfo {pages} {243602} (\bibinfo {year} {2002})}\BibitemShut {NoStop}%
\bibitem [{\citenamefont {Pöltl}\ \emph {et~al.}(2012)\citenamefont {Pöltl},
  \citenamefont {Emary},\ and\ \citenamefont {Brandes}}]{Poltl2012}%
  \BibitemOpen
  \bibfield  {author} {\bibinfo {author} {\bibfnamefont {C.}~\bibnamefont
  {Pöltl}}, \bibinfo {author} {\bibfnamefont {C.}~\bibnamefont {Emary}},\ and\
  \bibinfo {author} {\bibfnamefont {T.}~\bibnamefont {Brandes}},\ }\bibfield
  {title} {\bibinfo {title} {Spin entangled two-particle dark state in quantum
  transport through coupled quantum dots},\ }\href
  {https://doi.org/10.1103/PhysRevB.87.045416} {\bibfield  {journal} {\bibinfo
  {journal} {Physical Review B}\ }\textbf {\bibinfo {volume} {87}} (\bibinfo
  {year} {2012})}\BibitemShut {NoStop}%
\bibitem [{\citenamefont {Tanamoto}\ \emph {et~al.}(2012)\citenamefont
  {Tanamoto}, \citenamefont {Ono},\ and\ \citenamefont {Nori}}]{Tanamoto2012}%
  \BibitemOpen
  \bibfield  {author} {\bibinfo {author} {\bibfnamefont {T.}~\bibnamefont
  {Tanamoto}}, \bibinfo {author} {\bibfnamefont {K.}~\bibnamefont {Ono}},\ and\
  \bibinfo {author} {\bibfnamefont {F.}~\bibnamefont {Nori}},\ }\bibfield
  {title} {\bibinfo {title} {Steady-state solution for dark states using a
  three-level system in coupled quantum dots},\ }\href
  {https://doi.org/10.7567/JJAP.51.02BJ07} {\bibfield  {journal} {\bibinfo
  {journal} {Japanese Journal of Applied Physics}\ }\textbf {\bibinfo {volume}
  {51}},\ \bibinfo {pages} {02BJ07} (\bibinfo {year} {2012})}\BibitemShut
  {NoStop}%
\bibitem [{\citenamefont {Hansom}\ \emph {et~al.}(2014)\citenamefont {Hansom},
  \citenamefont {Schulte}, \citenamefont {Le~Gall}, \citenamefont {Matthiesen},
  \citenamefont {Clarke}, \citenamefont {Hugues}, \citenamefont {Taylor},\ and\
  \citenamefont {Atatüre}}]{Hansom2014}%
  \BibitemOpen
  \bibfield  {author} {\bibinfo {author} {\bibfnamefont {J.}~\bibnamefont
  {Hansom}}, \bibinfo {author} {\bibfnamefont {C.~H.~H.}\ \bibnamefont
  {Schulte}}, \bibinfo {author} {\bibfnamefont {C.}~\bibnamefont {Le~Gall}},
  \bibinfo {author} {\bibfnamefont {C.}~\bibnamefont {Matthiesen}}, \bibinfo
  {author} {\bibfnamefont {E.}~\bibnamefont {Clarke}}, \bibinfo {author}
  {\bibfnamefont {M.}~\bibnamefont {Hugues}}, \bibinfo {author} {\bibfnamefont
  {J.~M.}\ \bibnamefont {Taylor}},\ and\ \bibinfo {author} {\bibfnamefont
  {M.}~\bibnamefont {Atatüre}},\ }\bibfield  {title} {\bibinfo {title}
  {Environment-assisted quantum control of a solid-state spin via coherent dark
  states},\ }\href {https://doi.org/10.1038/nphys3077} {\bibfield  {journal}
  {\bibinfo  {journal} {Nature Physics}\ }\textbf {\bibinfo {volume} {10}},\
  \bibinfo {pages} {725} (\bibinfo {year} {2014})}\BibitemShut {NoStop}%
\bibitem [{\citenamefont {Kozyrev}\ and\ \citenamefont
  {Volovich}(2018)}]{Kozyrev2016}%
  \BibitemOpen
  \bibfield  {author} {\bibinfo {author} {\bibfnamefont {S.~V.}\ \bibnamefont
  {Kozyrev}}\ and\ \bibinfo {author} {\bibfnamefont {I.~V.}\ \bibnamefont
  {Volovich}},\ }\bibinfo {title} {Dark states in quantum photosynthesis},\ in\
  \href {https://doi.org/10.1007/978-3-319-91092-5_2} {\emph {\bibinfo
  {booktitle} {Trends in Biomathematics: Modeling, Optimization and
  Computational Problems: Selected works from the BIOMAT Consortium Lectures,
  Moscow 2017}}},\ \bibinfo {editor} {edited by\ \bibinfo {editor}
  {\bibfnamefont {R.~P.}\ \bibnamefont {Mondaini}}}\ (\bibinfo  {publisher}
  {Springer International Publishing},\ \bibinfo {address} {Cham},\ \bibinfo
  {year} {2018})\ pp.\ \bibinfo {pages} {13--26}\BibitemShut {NoStop}%
\bibitem [{\citenamefont {Kulagin}\ and\ \citenamefont
  {Ozhigov}(2020)}]{Ozhigov2020}%
  \BibitemOpen
  \bibfield  {author} {\bibinfo {author} {\bibfnamefont {A.~V.}\ \bibnamefont
  {Kulagin}}\ and\ \bibinfo {author} {\bibfnamefont {Y.~I.}\ \bibnamefont
  {Ozhigov}},\ }\bibfield  {title} {\bibinfo {title} {Optical selection of dark
  states of multilevel atomic ensembles},\ }\href
  {https://doi.org/10.1007/s10598-021-09504-3} {\bibfield  {journal} {\bibinfo
  {journal} {Computational Mathematics and Modeling}\ }\textbf {\bibinfo
  {volume} {31}},\ \bibinfo {pages} {431} (\bibinfo {year} {2020})}\BibitemShut
  {NoStop}%
\bibitem [{\citenamefont {Afanasyev}\ \emph {et~al.}(2022)\citenamefont
  {Afanasyev}, \citenamefont {Chen}, \citenamefont {Ozhigov},\ and\
  \citenamefont {You}}]{Afanasyev2022}%
  \BibitemOpen
  \bibfield  {author} {\bibinfo {author} {\bibfnamefont {V.}~\bibnamefont
  {Afanasyev}}, \bibinfo {author} {\bibfnamefont {R.}~\bibnamefont {Chen}},
  \bibinfo {author} {\bibfnamefont {Y.}~\bibnamefont {Ozhigov}},\ and\ \bibinfo
  {author} {\bibfnamefont {J.}~\bibnamefont {You}},\ }\bibfield  {title}
  {\bibinfo {title} {Collapse of dark states in {T}avis--{C}ummings model},\
  }\href {https://doi.org/10.1007/s10598-023-09571-8} {\bibfield  {journal}
  {\bibinfo  {journal} {Comput Math Model}\ }\textbf {\bibinfo {volume} {33}},\
  \bibinfo {pages} {273} (\bibinfo {year} {2022})}\BibitemShut {NoStop}%
\bibitem [{\citenamefont {Prasad}\ and\ \citenamefont
  {Martin}(2018)}]{Prasad2018}%
  \BibitemOpen
  \bibfield  {author} {\bibinfo {author} {\bibfnamefont {S.}~\bibnamefont
  {Prasad}}\ and\ \bibinfo {author} {\bibfnamefont {A.}~\bibnamefont
  {Martin}},\ }\bibfield  {title} {\bibinfo {title} {Effective three-body
  interactions in {J}aynes--{C}ummings--{H}ubbard systems},\ }\href
  {https://doi.org/10.1038/s41598-018-33907-9} {\bibfield  {journal} {\bibinfo
  {journal} {Sci Rep}\ }\textbf {\bibinfo {volume} {8}},\ \bibinfo {pages}
  {16253} (\bibinfo {year} {2018})}\BibitemShut {NoStop}%
\bibitem [{\citenamefont {Wei}\ \emph {et~al.}(2021)\citenamefont {Wei},
  \citenamefont {Zhang}, \citenamefont {Greschner}, \citenamefont {Scott},\
  and\ \citenamefont {Zhang}}]{Wei2021}%
  \BibitemOpen
  \bibfield  {author} {\bibinfo {author} {\bibfnamefont {H.}~\bibnamefont
  {Wei}}, \bibinfo {author} {\bibfnamefont {J.}~\bibnamefont {Zhang}}, \bibinfo
  {author} {\bibfnamefont {S.}~\bibnamefont {Greschner}}, \bibinfo {author}
  {\bibfnamefont {T.~C.}\ \bibnamefont {Scott}},\ and\ \bibinfo {author}
  {\bibfnamefont {W.}~\bibnamefont {Zhang}},\ }\bibfield  {title} {\bibinfo
  {title} {Quantum monte carlo study of superradiant supersolid of light in the
  extended {J}aynes--{C}ummings--{H}ubbard model},\ }\href
  {https://doi.org/10.1103/PhysRevB.103.184501} {\bibfield  {journal} {\bibinfo
   {journal} {Phys. Rev. B}\ }\textbf {\bibinfo {volume} {103}},\ \bibinfo
  {pages} {184501} (\bibinfo {year} {2021})}\BibitemShut {NoStop}%
\bibitem [{\citenamefont {Guo}\ \emph {et~al.}(2019)\citenamefont {Guo},
  \citenamefont {Greschner}, \citenamefont {Zhu},\ and\ \citenamefont
  {Zhang}}]{Guo2019}%
  \BibitemOpen
  \bibfield  {author} {\bibinfo {author} {\bibfnamefont {L.}~\bibnamefont
  {Guo}}, \bibinfo {author} {\bibfnamefont {S.}~\bibnamefont {Greschner}},
  \bibinfo {author} {\bibfnamefont {S.}~\bibnamefont {Zhu}},\ and\ \bibinfo
  {author} {\bibfnamefont {W.}~\bibnamefont {Zhang}},\ }\bibfield  {title}
  {\bibinfo {title} {Supersolid and pair correlations of the extended
  {J}aynes--{C}ummings--{H}ubbard model on triangular lattices},\ }\href
  {https://doi.org/10.1103/PhysRevA.100.033614} {\bibfield  {journal} {\bibinfo
   {journal} {Phys. Rev. A}\ }\textbf {\bibinfo {volume} {100}},\ \bibinfo
  {pages} {033614} (\bibinfo {year} {2019})}\BibitemShut {NoStop}%
\bibitem [{\citenamefont {Victorova}\ \emph {et~al.}(2020)\citenamefont
  {Victorova}, \citenamefont {Kulagin},\ and\ \citenamefont
  {Ozhigov}}]{Victorova2020}%
  \BibitemOpen
  \bibfield  {author} {\bibinfo {author} {\bibfnamefont {N.}~\bibnamefont
  {Victorova}}, \bibinfo {author} {\bibfnamefont {A.}~\bibnamefont {Kulagin}},\
  and\ \bibinfo {author} {\bibfnamefont {Y.}~\bibnamefont {Ozhigov}},\
  }\bibfield  {title} {\bibinfo {title} {Quasi-classical description of the
  “quantum bottleneck” effect for thermal relaxation of an atom in a
  resonator},\ }\href {https://doi.org/10.1007/s10598-020-09470-2} {\bibfield
  {journal} {\bibinfo  {journal} {Comput Math Model}\ }\textbf {\bibinfo
  {volume} {31}},\ \bibinfo {pages} {1} (\bibinfo {year} {2020})}\BibitemShut
  {NoStop}%
\bibitem [{\citenamefont {Afanasyev}\ \emph {et~al.}(2021)\citenamefont
  {Afanasyev}, \citenamefont {Zheng}, \citenamefont {Kulagin}, \citenamefont
  {Miao}, \citenamefont {Ozhigov}, \citenamefont {Li},\ and\ \citenamefont
  {Victorova}}]{Ozhigov2021}%
  \BibitemOpen
  \bibfield  {author} {\bibinfo {author} {\bibfnamefont {V.}~\bibnamefont
  {Afanasyev}}, \bibinfo {author} {\bibfnamefont {K.}~\bibnamefont {Zheng}},
  \bibinfo {author} {\bibfnamefont {A.}~\bibnamefont {Kulagin}}, \bibinfo
  {author} {\bibfnamefont {H.-h.}\ \bibnamefont {Miao}}, \bibinfo {author}
  {\bibfnamefont {Y.}~\bibnamefont {Ozhigov}}, \bibinfo {author} {\bibfnamefont
  {W.}~\bibnamefont {Li}},\ and\ \bibinfo {author} {\bibfnamefont
  {N.}~\bibnamefont {Victorova}},\ }\bibfield  {title} {\bibinfo {title} {About
  chemical modifications of finite dimensional {QED} models},\ }\href
  {https://doi.org/10.33581/1561-4085-2021-24-3-230-241} {\bibfield  {journal}
  {\bibinfo  {journal} {Nonlinear Phenomena in Complex Systems}\ }\textbf
  {\bibinfo {volume} {24}},\ \bibinfo {pages} {230} (\bibinfo {year}
  {2021})}\BibitemShut {NoStop}%
\bibitem [{\citenamefont {Kulagin}\ and\ \citenamefont
  {Ozhigov}(2022)}]{Kulagin2022}%
  \BibitemOpen
  \bibfield  {author} {\bibinfo {author} {\bibfnamefont {A.}~\bibnamefont
  {Kulagin}}\ and\ \bibinfo {author} {\bibfnamefont {Y.}~\bibnamefont
  {Ozhigov}},\ }\bibfield  {title} {\bibinfo {title} {Realization of grover
  search algorithm on the optical cavities},\ }\href
  {https://doi.org/10.1134/S1995080222070162} {\bibfield  {journal} {\bibinfo
  {journal} {Lobachevskii J Math}\ }\textbf {\bibinfo {volume} {43}},\ \bibinfo
  {pages} {864} (\bibinfo {year} {2022})}\BibitemShut {NoStop}%
\bibitem [{\citenamefont {Ozhigov}\ and\ \citenamefont
  {Pluzhnikov}(2022)}]{Pluzhnikov2022}%
  \BibitemOpen
  \bibfield  {author} {\bibinfo {author} {\bibfnamefont {Y.}~\bibnamefont
  {Ozhigov}}\ and\ \bibinfo {author} {\bibfnamefont {I.}~\bibnamefont
  {Pluzhnikov}},\ }\bibfield  {title} {\bibinfo {title} {Superimposition and
  antagonism in chain synthesis using entangled biphotonic control},\ }\href
  {https://doi.org/10.1007/s10598-022-09553-2} {\bibfield  {journal} {\bibinfo
  {journal} {Comput Math Model}\ }\textbf {\bibinfo {volume} {33}},\ \bibinfo
  {pages} {24} (\bibinfo {year} {2022})}\BibitemShut {NoStop}%
\bibitem [{\citenamefont {Chen}\ \emph {et~al.}(2022)\citenamefont {Chen},
  \citenamefont {Ozhigov},\ and\ \citenamefont {You}}]{Chen2022}%
  \BibitemOpen
  \bibfield  {author} {\bibinfo {author} {\bibfnamefont {R.}~\bibnamefont
  {Chen}}, \bibinfo {author} {\bibfnamefont {Y.~I.}\ \bibnamefont {Ozhigov}},\
  and\ \bibinfo {author} {\bibfnamefont {J.~C.}\ \bibnamefont {You}},\
  }\bibfield  {title} {\bibinfo {title} {Qualitative model of the hydrogen
  peroxide positive ion in a heat bath},\ }\href
  {https://doi.org/10.1007/s10598-023-09583-4} {\bibfield  {journal} {\bibinfo
  {journal} {Comput Math Model}\ }\textbf {\bibinfo {volume} {33}},\ \bibinfo
  {pages} {408} (\bibinfo {year} {2022})}\BibitemShut {NoStop}%
\bibitem [{\citenamefont {hui Miao}\ and\ \citenamefont
  {Ozhigov}(2023)}]{Miao2023}%
  \BibitemOpen
  \bibfield  {author} {\bibinfo {author} {\bibfnamefont {H.}~\bibnamefont {hui
  Miao}}\ and\ \bibinfo {author} {\bibfnamefont {Y.~I.}\ \bibnamefont
  {Ozhigov}},\ }\bibfield  {title} {\bibinfo {title} {Using a modified version
  of the {T}avis-{C}ummings-{H}ubbard model to simulate the formation of
  neutral hydrogen molecule},\ }\href
  {https://doi.org/https://doi.org/10.1016/j.physa.2023.128851} {\bibfield
  {journal} {\bibinfo  {journal} {Physica A: Statistical Mechanics and its
  Applications}\ }\textbf {\bibinfo {volume} {622}},\ \bibinfo {pages} {128851}
  (\bibinfo {year} {2023})}\BibitemShut {NoStop}%
\bibitem [{\citenamefont {Miao}\ and\ \citenamefont
  {Ozhigov}(2023)}]{MiaoOzhigov2023}%
  \BibitemOpen
  \bibfield  {author} {\bibinfo {author} {\bibfnamefont {H.-h.}\ \bibnamefont
  {Miao}}\ and\ \bibinfo {author} {\bibfnamefont {Y.~I.}\ \bibnamefont
  {Ozhigov}},\ }\bibfield  {title} {\bibinfo {title} {Comparing the effects of
  nuclear and electron spins on the formation of neutral hydrogen molecule},\
  }\href {https://doi.org/10.1134/S1995080223080401} {\bibfield  {journal}
  {\bibinfo  {journal} {Lobachevskii Journal of Mathematics}\ }\textbf
  {\bibinfo {volume} {44}},\ \bibinfo {pages} {3111} (\bibinfo {year}
  {2023})}\BibitemShut {NoStop}%
\bibitem [{\citenamefont {Li}\ \emph {et~al.}(2024)\citenamefont {Li},
  \citenamefont {Miao},\ and\ \citenamefont {Ozhigov}}]{LiMiao2024}%
  \BibitemOpen
  \bibfield  {author} {\bibinfo {author} {\bibfnamefont {W.}~\bibnamefont
  {Li}}, \bibinfo {author} {\bibfnamefont {H.-h.}\ \bibnamefont {Miao}},\ and\
  \bibinfo {author} {\bibfnamefont {Y.~I.}\ \bibnamefont {Ozhigov}},\
  }\bibfield  {title} {\bibinfo {title} {Supercomputer model of
  finite-dimensional quantum electrodynamics applications},\ }\href
  {https://doi.org/10.1134/S1995080224603849} {\bibfield  {journal} {\bibinfo
  {journal} {Lobachevskii Journal of Mathematics}\ }\textbf {\bibinfo {volume}
  {45}},\ \bibinfo {pages} {3097} (\bibinfo {year} {2024})}\BibitemShut
  {NoStop}%
\bibitem [{\citenamefont {Miao}\ and\ \citenamefont
  {Ozhigov}(2024)}]{MiaoOzhigov2024}%
  \BibitemOpen
  \bibfield  {author} {\bibinfo {author} {\bibfnamefont {H.-h.}\ \bibnamefont
  {Miao}}\ and\ \bibinfo {author} {\bibfnamefont {Y.~I.}\ \bibnamefont
  {Ozhigov}},\ }\bibfield  {title} {\bibinfo {title} {Distributed computing
  quantum unitary evolution},\ }\href
  {https://doi.org/10.1134/S1995080224603904} {\bibfield  {journal} {\bibinfo
  {journal} {Lobachevskii Journal of Mathematics}\ }\textbf {\bibinfo {volume}
  {45}},\ \bibinfo {pages} {3121} (\bibinfo {year} {2024})}\BibitemShut
  {NoStop}%
\bibitem [{\citenamefont {Huelga}\ and\ \citenamefont
  {Plenio}(2013)}]{Huelga2013}%
  \BibitemOpen
  \bibfield  {author} {\bibinfo {author} {\bibfnamefont {S.}~\bibnamefont
  {Huelga}}\ and\ \bibinfo {author} {\bibfnamefont {M.}~\bibnamefont
  {Plenio}},\ }\bibfield  {title} {\bibinfo {title} {Vibrations, quanta and
  biology},\ }\href {https://doi.org/10.1080/00405000.2013.829687} {\bibfield
  {journal} {\bibinfo  {journal} {Contemporary Physics}\ }\textbf {\bibinfo
  {volume} {54}},\ \bibinfo {pages} {181} (\bibinfo {year} {2013})}\BibitemShut
  {NoStop}%
\bibitem [{\citenamefont {Kulagin}\ \emph {et~al.}(2019)\citenamefont
  {Kulagin}, \citenamefont {Ladunov}, \citenamefont {Ozhigov}, \citenamefont
  {Skovoroda},\ and\ \citenamefont {Victorova}}]{Kulagin2019}%
  \BibitemOpen
  \bibfield  {author} {\bibinfo {author} {\bibfnamefont {A.~V.}\ \bibnamefont
  {Kulagin}}, \bibinfo {author} {\bibfnamefont {V.~Y.}\ \bibnamefont
  {Ladunov}}, \bibinfo {author} {\bibfnamefont {Y.~I.}\ \bibnamefont
  {Ozhigov}}, \bibinfo {author} {\bibfnamefont {N.~A.}\ \bibnamefont
  {Skovoroda}},\ and\ \bibinfo {author} {\bibfnamefont {N.~B.}\ \bibnamefont
  {Victorova}},\ }\bibfield  {title} {\bibinfo {title} {Homogeneous atomic
  ensembles and single-mode field: review of simulation results},\ }in\ \href
  {https://doi.org/10.1117/12.2521763} {\emph {\bibinfo {booktitle}
  {International Conference on Micro- and Nano-Electronics 2018}}},\ Vol.\
  \bibinfo {volume} {11022},\ \bibinfo {editor} {edited by\ \bibinfo {editor}
  {\bibfnamefont {V.~F.}\ \bibnamefont {Lukichev}}\ and\ \bibinfo {editor}
  {\bibfnamefont {K.~V.}\ \bibnamefont {Rudenko}}},\ \bibinfo {organization}
  {International Society for Optics and Photonics}\ (\bibinfo  {publisher}
  {SPIE},\ \bibinfo {year} {2019})\ p.\ \bibinfo {pages} {110222C}\BibitemShut
  {NoStop}%
\bibitem [{\citenamefont {Fecko}\ \emph {et~al.}(2003)\citenamefont {Fecko},
  \citenamefont {Eaves}, \citenamefont {Loparo}, \citenamefont {Tokmakoff},\
  and\ \citenamefont {Geissler}}]{Fecko2003}%
  \BibitemOpen
  \bibfield  {author} {\bibinfo {author} {\bibfnamefont {C.~J.}\ \bibnamefont
  {Fecko}}, \bibinfo {author} {\bibfnamefont {J.~D.}\ \bibnamefont {Eaves}},
  \bibinfo {author} {\bibfnamefont {J.~J.}\ \bibnamefont {Loparo}}, \bibinfo
  {author} {\bibfnamefont {A.}~\bibnamefont {Tokmakoff}},\ and\ \bibinfo
  {author} {\bibfnamefont {P.~L.}\ \bibnamefont {Geissler}},\ }\bibfield
  {title} {\bibinfo {title} {Ultrafast hydrogen-bond dynamics in the infrared
  spectroscopy of water},\ }\href {https://doi.org/10.1126/science.1087251}
  {\bibfield  {journal} {\bibinfo  {journal} {Science}\ }\textbf {\bibinfo
  {volume} {301}},\ \bibinfo {pages} {1698} (\bibinfo {year}
  {2003})}\BibitemShut {NoStop}%
\bibitem [{\citenamefont {Lawrence}\ and\ \citenamefont
  {Skinner}(2003)}]{Lawrence2003}%
  \BibitemOpen
  \bibfield  {author} {\bibinfo {author} {\bibfnamefont {C.}~\bibnamefont
  {Lawrence}}\ and\ \bibinfo {author} {\bibfnamefont {J.}~\bibnamefont
  {Skinner}},\ }\bibfield  {title} {\bibinfo {title} {Ultrafast infrared
  spectroscopy probes hydrogen-bonding dynamics in liquid water},\ }\href
  {https://doi.org/https://doi.org/10.1016/S0009-2614(02)02039-0} {\bibfield
  {journal} {\bibinfo  {journal} {Chemical Physics Letters}\ }\textbf {\bibinfo
  {volume} {369}},\ \bibinfo {pages} {472} (\bibinfo {year}
  {2003})}\BibitemShut {NoStop}%
\bibitem [{\citenamefont {Møller}\ \emph {et~al.}(2004)\citenamefont
  {Møller}, \citenamefont {Rey},\ and\ \citenamefont {Hynes}}]{Moller2004}%
  \BibitemOpen
  \bibfield  {author} {\bibinfo {author} {\bibfnamefont {K.~B.}\ \bibnamefont
  {Møller}}, \bibinfo {author} {\bibfnamefont {R.}~\bibnamefont {Rey}},\ and\
  \bibinfo {author} {\bibfnamefont {J.~T.}\ \bibnamefont {Hynes}},\ }\bibfield
  {title} {\bibinfo {title} {Hydrogen bond dynamics in water and ultrafast
  infrared spectroscopy: A theoretical study},\ }\href
  {https://doi.org/10.1021/jp035935r} {\bibfield  {journal} {\bibinfo
  {journal} {The Journal of Physical Chemistry A}\ }\textbf {\bibinfo {volume}
  {108}},\ \bibinfo {pages} {1275} (\bibinfo {year} {2004})}\BibitemShut
  {NoStop}%
\bibitem [{\citenamefont {Wu}\ and\ \citenamefont {Yang}(2007)}]{Wu2007}%
  \BibitemOpen
  \bibfield  {author} {\bibinfo {author} {\bibfnamefont {Y.}~\bibnamefont
  {Wu}}\ and\ \bibinfo {author} {\bibfnamefont {X.}~\bibnamefont {Yang}},\
  }\bibfield  {title} {\bibinfo {title} {Strong-coupling theory of periodically
  driven two-level systems},\ }\href
  {https://doi.org/10.1103/PhysRevLett.98.013601} {\bibfield  {journal}
  {\bibinfo  {journal} {Phys. Rev. Lett.}\ }\textbf {\bibinfo {volume} {98}},\
  \bibinfo {pages} {013601} (\bibinfo {year} {2007})}\BibitemShut {NoStop}%
\bibitem [{\citenamefont {Dum}\ \emph {et~al.}(1992)\citenamefont {Dum},
  \citenamefont {Zoller},\ and\ \citenamefont {Ritsch}}]{Dum1992}%
  \BibitemOpen
  \bibfield  {author} {\bibinfo {author} {\bibfnamefont {R.}~\bibnamefont
  {Dum}}, \bibinfo {author} {\bibfnamefont {P.}~\bibnamefont {Zoller}},\ and\
  \bibinfo {author} {\bibfnamefont {H.}~\bibnamefont {Ritsch}},\ }\bibfield
  {title} {\bibinfo {title} {Monte carlo simulation of the atomic master
  equation for spontaneous emission},\ }\href
  {https://doi.org/10.1103/PhysRevA.45.4879} {\bibfield  {journal} {\bibinfo
  {journal} {Phys. Rev. A}\ }\textbf {\bibinfo {volume} {45}},\ \bibinfo
  {pages} {4879} (\bibinfo {year} {1992})}\BibitemShut {NoStop}%
\bibitem [{\citenamefont {Ozhigov}\ and\ \citenamefont {You}(2023)}]{You2023}%
  \BibitemOpen
  \bibfield  {author} {\bibinfo {author} {\bibfnamefont {Y.~I.}\ \bibnamefont
  {Ozhigov}}\ and\ \bibinfo {author} {\bibfnamefont {J.~C.}\ \bibnamefont
  {You}},\ }\bibfield  {title} {\bibinfo {title} {Description of the
  non-markovian dynamics of atoms in terms of a pure state},\ }\href
  {https://doi.org/10.1007/s10598-024-09596-7} {\bibfield  {journal} {\bibinfo
  {journal} {Comput Math Model}\ }\textbf {\bibinfo {volume} {34}},\ \bibinfo
  {pages} {75} (\bibinfo {year} {2023})}\BibitemShut {NoStop}%
\bibitem [{\citenamefont {Hennrich}\ \emph {et~al.}(2000)\citenamefont
  {Hennrich}, \citenamefont {Legero}, \citenamefont {Kuhn},\ and\ \citenamefont
  {Rempe}}]{Hennrich2000}%
  \BibitemOpen
  \bibfield  {author} {\bibinfo {author} {\bibfnamefont {M.}~\bibnamefont
  {Hennrich}}, \bibinfo {author} {\bibfnamefont {T.}~\bibnamefont {Legero}},
  \bibinfo {author} {\bibfnamefont {A.}~\bibnamefont {Kuhn}},\ and\ \bibinfo
  {author} {\bibfnamefont {G.}~\bibnamefont {Rempe}},\ }\bibfield  {title}
  {\bibinfo {title} {Vacuum-stimulated raman scattering based on adiabatic
  passage in a high-finesse optical cavity},\ }\href
  {https://doi.org/10.1103/PhysRevLett.85.4872} {\bibfield  {journal} {\bibinfo
   {journal} {Phys. Rev. Lett.}\ }\textbf {\bibinfo {volume} {85}},\ \bibinfo
  {pages} {4872} (\bibinfo {year} {2000})}\BibitemShut {NoStop}%
\bibitem [{\citenamefont {Raimond}\ \emph {et~al.}(2001)\citenamefont
  {Raimond}, \citenamefont {Brune},\ and\ \citenamefont
  {Haroche}}]{Raimond2001}%
  \BibitemOpen
  \bibfield  {author} {\bibinfo {author} {\bibfnamefont {J.~M.}\ \bibnamefont
  {Raimond}}, \bibinfo {author} {\bibfnamefont {M.}~\bibnamefont {Brune}},\
  and\ \bibinfo {author} {\bibfnamefont {S.}~\bibnamefont {Haroche}},\
  }\bibfield  {title} {\bibinfo {title} {Manipulating quantum entanglement with
  atoms and photons in a cavity},\ }\href
  {https://doi.org/10.1103/RevModPhys.73.565} {\bibfield  {journal} {\bibinfo
  {journal} {Rev. Mod. Phys.}\ }\textbf {\bibinfo {volume} {73}},\ \bibinfo
  {pages} {565} (\bibinfo {year} {2001})}\BibitemShut {NoStop}%
\bibitem [{\citenamefont {Voevodin}\ \emph {et~al.}(2019)\citenamefont
  {Voevodin}, \citenamefont {Antonov}, \citenamefont {Nikitenko}, \citenamefont
  {Shvets}, \citenamefont {Sobolev}, \citenamefont {Sidorov}, \citenamefont
  {Stefanov}, \citenamefont {Voevodin},\ and\ \citenamefont
  {Zhumatiy}}]{Voevodin2019}%
  \BibitemOpen
  \bibfield  {author} {\bibinfo {author} {\bibfnamefont {V.}~\bibnamefont
  {Voevodin}}, \bibinfo {author} {\bibfnamefont {A.}~\bibnamefont {Antonov}},
  \bibinfo {author} {\bibfnamefont {D.}~\bibnamefont {Nikitenko}}, \bibinfo
  {author} {\bibfnamefont {P.}~\bibnamefont {Shvets}}, \bibinfo {author}
  {\bibfnamefont {S.}~\bibnamefont {Sobolev}}, \bibinfo {author} {\bibfnamefont
  {I.}~\bibnamefont {Sidorov}}, \bibinfo {author} {\bibfnamefont
  {K.}~\bibnamefont {Stefanov}}, \bibinfo {author} {\bibfnamefont
  {V.}~\bibnamefont {Voevodin}},\ and\ \bibinfo {author} {\bibfnamefont
  {S.}~\bibnamefont {Zhumatiy}},\ }\bibfield  {title} {\bibinfo {title}
  {Supercomputer {L}omonosov-2: Large scale, deep monitoring and fine analytics
  for the user community},\ }\href {https://doi.org/10.14529/jsfi190201}
  {\bibfield  {journal} {\bibinfo  {journal} {Supercomputing Frontiers and
  Innovations}\ }\textbf {\bibinfo {volume} {6}},\ \bibinfo {pages} {4}
  (\bibinfo {year} {2019})}\BibitemShut {NoStop}%
\bibitem [{\citenamefont {Fontaine-Vive}\ \emph {et~al.}(2006)\citenamefont
  {Fontaine-Vive}, \citenamefont {Johnson}, \citenamefont {Kearley},
  \citenamefont {Cowan},\ and\ \citenamefont {Howard}}]{Fontaine2006}%
  \BibitemOpen
  \bibfield  {author} {\bibinfo {author} {\bibfnamefont {F.}~\bibnamefont
  {Fontaine-Vive}}, \bibinfo {author} {\bibfnamefont {M.~R.}\ \bibnamefont
  {Johnson}}, \bibinfo {author} {\bibfnamefont {G.~J.}\ \bibnamefont
  {Kearley}}, \bibinfo {author} {\bibfnamefont {J.~A.}\ \bibnamefont {Cowan}},\
  and\ \bibinfo {author} {\bibfnamefont {J.~A.}\ \bibnamefont {Howard}},\
  }\bibfield  {title} {\bibinfo {title} {Phonon driven proton transfer in
  crystals with short strong hydrogen bonds},\ }\href
  {https://doi.org/10.1063/1.2206774} {\bibfield  {journal} {\bibinfo
  {journal} {The Journal of Chemical Physics}\ }\textbf {\bibinfo {volume}
  {124}},\ \bibinfo {pages} {234503} (\bibinfo {year} {2006})}\BibitemShut
  {NoStop}%
\end{thebibliography}%

\end{document}